\documentclass[conference]{IEEEtran}
\IEEEoverridecommandlockouts
\usepackage{cite}
\usepackage{amsmath,amssymb,amsfonts}
\usepackage{graphicx}
\usepackage{textcomp}
\usepackage{xcolor}
\def\BibTeX{{\rm B\kern-.05em{\sc i\kern-.025em b}\kern-.08em
    T\kern-.1667em\lower.7ex\hbox{E}\kern-.125emX}}
\usepackage{amssymb}
\usepackage{amsmath}
\usepackage{xcolor}
\usepackage{algorithm}%
\usepackage{algorithmicx}%
\usepackage{algpseudocode}%
\usepackage{array}
\usepackage{hyperref}
\usepackage{subfig}
\usepackage{comment}
\usepackage[dvipsnames]{xcolor}
\usepackage[normalem]{ulem}

\begin{document}




\title{On Enhancing Delay SLAs in TCP Networks through Joint Routing and Transport Assistant Deployment \thanks{This work has been published in the International Journal of Network Management. 
The final version is available at: \href{https://doi.org/10.1002/nem.70006}{https://doi.org/10.1002/nem.70006}}}

\author{
\IEEEauthorblockN{José Gómez-delaHiz\IEEEauthorrefmark{1},
Mohamed Faten Zhani\IEEEauthorrefmark{2},
Jaime Galán-Jiménez\IEEEauthorrefmark{3},
John Kaippallimalil\IEEEauthorrefmark{4}}
\IEEEauthorblockA{\IEEEauthorrefmark{1}\textit{Department of Computer Systems and Telematics Engineering}, University of Extremadura, Cáceres, Spain, jagomezdh@unex.es}
\IEEEauthorblockA{\IEEEauthorrefmark{2}\textit{Department of Networks and Multimedia}, ISITCom, University of Sousse, Sousse, Tunisia, mf.zhani@isitc.u-sousse.tn}
\IEEEauthorblockA{\IEEEauthorrefmark{3}\textit{Department of Computer Systems and Telematics Engineering}, University of Extremadura, Cáceres, Spain, jaime@unex.es}
\IEEEauthorblockA{\IEEEauthorrefmark{4}\textit{Futurewei Technologies, Inc.}, USA, john.kaippallimalil@futurewei.com}
}
\maketitle
\begin{abstract}
The Transport Control Protocol has long been the primary transport protocol for applications requiring performance and reliability over the Internet. Unfortunately, due its retransmission mechanism, TCP incurs high packet delivery delays when segments are lost. To address this issue, previous research proposed to use a novel network function, namely Transport Assistant, deployed within the network to cache and~retransmit lost packets, thus reducing retransmission delays. 
In this paper, we propose to jointly route the flows and deploy TAs in order to~minimize packet delivery delays in best-effort networks (scenario~1) or to~satisfy delay-based Service Level Agreements in~QoS-based networks (scenario~2). We hence formulate the joint routing and TA deployment problem as Integer Linear Program for the two scenarios and propose a heuristic solution for large-scale instances of the problem. Through extensive simulations, we demonstrate the benefits of performing joint routing flows and TA deployment in reducing packet delivery delays (up to 16.4\%) while minimizing deployment costs (up to 60.98\%).
\end{abstract}



\begin{IEEEkeywords}
\textcolor{black}{NFV, QoS, Service Level Agreement, TCP, Transport Assistant}



\end{IEEEkeywords}




\section{Introduction}



\textcolor{black}{The Transmission Control Protocol (TCP) has served as the primary transport protocol for applications requiring reliability since the inception of the Internet}. TCP offers a connection-oriented communication service across the Internet, facilitating reliable transport between two endpoints over a non-reliable infrastructure underpinned by the Internet Protocol (IP) \cite{rfc793}. The TCP protocol integrates several functions, including flow control, congestion control, and mechanisms for detecting and retransmitting lost packets.


Through the utilization of timers and duplicate acknowledgments (ACKs), TCP can detect lost packets and trigger retransmission accordingly. However, this mechanism results in significant delays for the delivery of packets especially when packets are lost\cite{Tao2023, Yahyaoui2023}. Specifically, the sender must wait for a timeout, approximately equivalent to a Round-Trip Time (RTT), before retransmitting the lost packet. Consequently, the packet delivery delay is at least three times the end-to-end source-to-destination delay. Such high delays can substantially impair the performance of various time-sensitive applications \cite{Hamad2023, Liu2023, Cao2023}.

A potential solution to mitigate this problem and reduce such delays would be to retransmit the lost packet from an intermediate node along the path between the source and the destination. As such, the retransmission is initiated from a node closer to the destination, and thereby reducing the retransmission delay. This approach is partially explored~in~\cite{zhani2020flexngia,Zhani-WMNC22,Yahyaoui-NOMS22} where the authors propose the deployment of a special network function named \textit{Transport Assistant} (TA). The TA is a network function that can be instantiated in an intermediate node to cache packets, detect losses, and retransmit lost packets. As it is located closer to the destination, the TA can promptly detect and retransmit lost packets, thereby reducing retransmission delays and obviating the need for the source node to resend the packet. 

\color{black}
Our previous efforts mainly focused on evaluating the impact of using~TAs on network performance, particularly on the end-to-end packet delivery delay, and also on studying the effect of its placement along the route between the source and the destination~\cite{JaimeTAPlacementNOMS2024, LuisNOMS2023}. However, in those works, we assumed that the flow route was already fixed and that TAs could be instantiated afterwards in one of~the~intermediate nodes to reduce the packet delivery delay. In contrast, the present work considers a more general and practical setting where the operator jointly decides on both the flow routing and the TA placement. In particular, this paper proposes the~following new contributions:
\begin{itemize}
    \item We first focus on~a~first scenario where we propose a~solution to jointly route TCP flows and select the location of~the~TAs within the route in order to minimize average flow packet delivery delays. This scenario assumes a~best-effort network, where the goal is to reduce end-to-end delays as~much~as~possible without providing any Service Level Agreement (SLA) guarantees.
    \item We also consider a~second scenario considering a~QoS-based network in which each flow is associated with an~SLA, defined as~an~upper bound on its packet delivery delay. In this case, the objective is to efficiently route flows and place TAs (when needed) with the goal of minimizing the average packet delivery delay across all flows, minimizing the deployment costs of the~TA, and reducing the penalty incurred when the~SLA~is~not~met. 
    \item We formulate the joint problem of flow routing and~TA~placement for the two studied scenarios as~Integer Linear Programs (ILP1 and ILP2). 
    \item As the flow routing and TA placement problems are NP-hard and could not be solved efficiently for large-scale instances, we propose using a~heuristic solution. Through extensive simulations, we show the~efficiency of this solution to reduce the average packet delivery delay in~best-effort networks and also satisfy delay-based SLAs with minimal costs for QoS-based networks.
 \end{itemize}
\color{black}

The rest of the paper is structured as follows.
Section~\ref{sec:BackgoundMaterial} provides relevant background material from our previous work~\cite{JaimeTAPlacementNOMS2024} to present the role and operation of TA as well as the mathematical model that will be used to~estimate the average packet delivery delay when a~TA is deployed.
Section~\ref{sec:problemDescription} provides an in-depth description of the addressed problem and the two studied scenarios. 
Section~\ref{sec:problemFormulation} presents the mathematical formulation of the problem for the two scenarios. 
Section~\ref{sec:heuristics} describes a heuristic algorithm that can solve the problem in~tractable times for both scenarios. 
%
Experimental evaluation is carried out in~Section~\ref{sec:results} to show the effectiveness of the proposed solutions.
A review of related works is provided in~Section~\ref{sec:RelatedWork}, and~finally Section~\ref{sec:conclusion} concludes the work.


\section{Background} \label{sec:BackgoundMaterial}

\textcolor{black}{Before presenting the~proposed solutions, we first review the role and deployment of~TAs, discuss representative implementations, and explain through a~mathematical model why their placement within the network is a critical factor influencing end-to-end packet delivery performance.}\\

\begin{figure}[t]
        \centering
         \includegraphics[width=0.8\columnwidth]{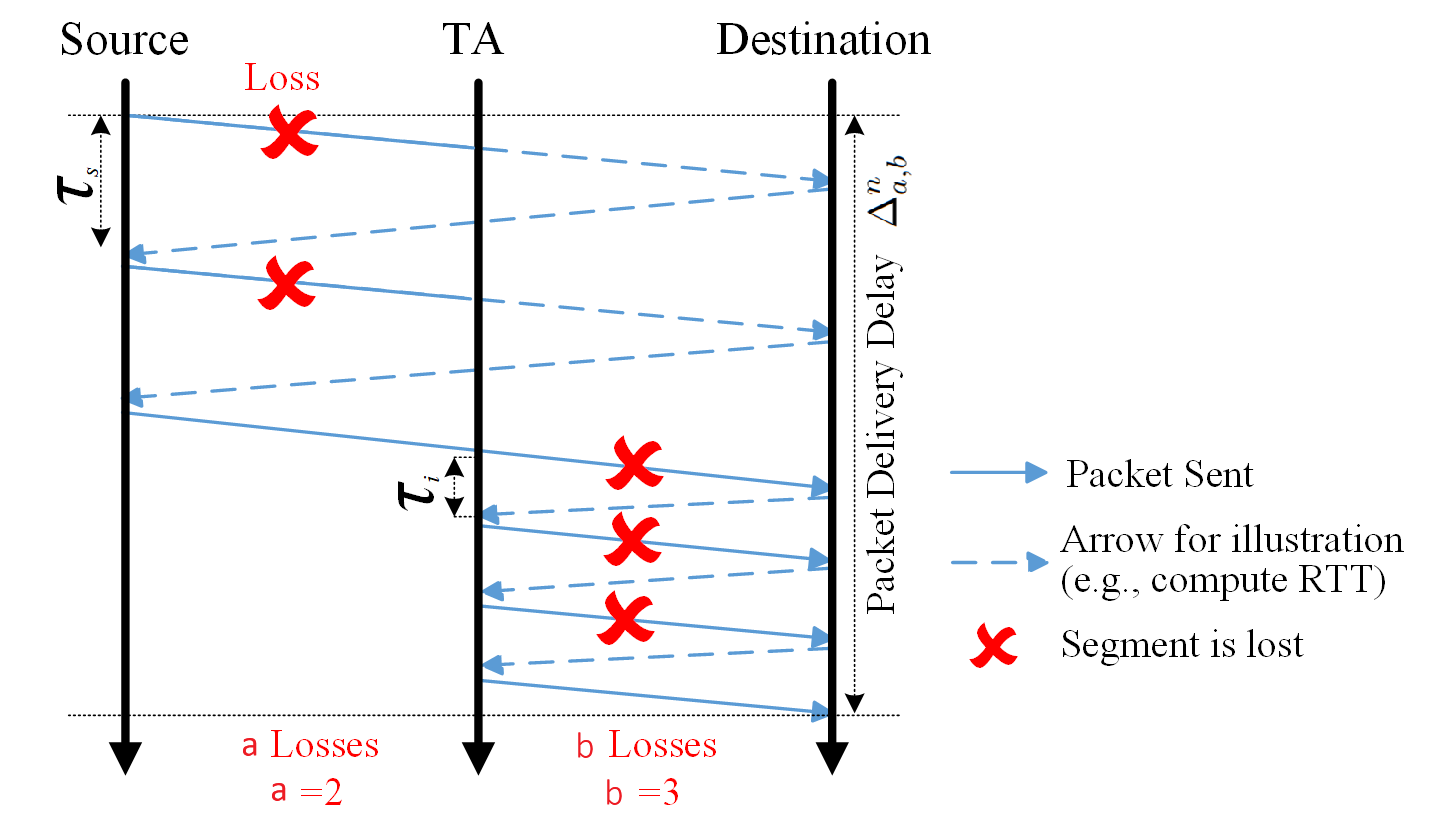}
         \caption{Packet delivery delay when the packet is lost $a$~times between
         the source and the TA, and $b$~times between the TA and the destination~\cite{JaimeTAPlacementNOMS2024}.}
        \label{fig:PacketDelivery}
\end{figure}

\textcolor{black}{
\textbf{- Transport Assistant Role and Deployment:} a~TA~is a~specialized network function that caches flow packets, detects losses before the source, and retransmits them in~order to reduce the overall packet delivery delay, as illustrated in~Fig.~\ref{fig:PacketDelivery}. In practice, a TA can be instantiated as a virtual network function in edge clouds or operator-controlled middleboxes, as advocated by recent proposals~\cite{LuizelliIM2015, zhani2020flexngia, moufakirITU2022} that support the deployment of service function chains including network functions along the path from the source to the destination.}
\textcolor{black}{Different implementations of TAs are already available. For~example, the one proposed in~\cite{Zhani-WMNC22} provides full transparency and compatibility with existing TCP stacks, without requiring any modification to TCP. Other solutions (e.g.,~\cite{Yahyaoui-NOMS22}), however, assume a modified TCP where the endpoints are aware of the TA and explicitly coordinate with it.}

\textcolor{black}{Regardless of the TA implementation, one of the key challenges pertaining to the deployment of TAs is how to select the optimal placement of the TA in the infrastructure. Indeed, the TA should be instantiated in one of the nodes within the path between the source and the destination. This placement will significantly impact the resulting the~performance of the packet delivery delay in the presence of the TA.} 
In our previous work~\cite{JaimeTAPlacementNOMS2024}, we developed a model based on probability theory to estimate the average packet delivery delay when the TA is deployed. In~the~following, we only provide the key elements of~this~model as it is used in this work to estimate the packet delivery delays and~guide the flow routing and~the~TA placement. \\

%

\textbf{- Modeling Packet Delivery Delay in the Presence of~the~TA:}  According to~\cite{JaimeTAPlacementNOMS2024}, the following equation provides the Expected Packet Delivery Delay (EPDD) for a flow $f$, when it is routed through a~path $p$ and assuming a~TA is~deployed in~node~$n \in p$ (see~Fig.~\ref{fig:PacketDelivery}):
%
\begin{equation}\label{eq:expDelivDelay}
\delta_{f, p}^n = \sum_{(a,b)\in \mathbb{N}^2} P_{f,p}^n(a,b) \Delta^n_{a,b}
\end{equation}
%
%
where $\Delta^n_{a,b}$ is the packet delivery time when the packet has been lost $a$~times between the source node of the flow and~the~TA (located at~node~$n$) and~then~lost $b$ times between the~TA and~the~flow destination~(Fig.~\ref{fig:PacketDelivery}). 
We~denote by~$P^n_{f,p}(a,b)$ the probability to have $a$ packets lost from flow $f$ between the source and the TA and $b$ packets lost between the TA and the destination, assuming the TA placed in node $n$ (see~Fig.~\ref{fig:PacketDelivery}). More~details on how the parameters of~the~equation are computed could be found~in~\cite{JaimeTAPlacementNOMS2024}. 

This equation is highly relevant as it captures the effect of TA placement on TCP packet delivery delay, while also accounting for potential packet loss and retransmissions. It~therefore forms the foundation of this work’s contribution, which focuses on jointly routing traffic flows and determining the optimal TA location in~the~network. In particular, the~equation will be used to predict packet delivery delays in our formulation and in the proposed heuristic.

\begin{figure}[t]
    \centering
    \includegraphics[width=0.9\linewidth]{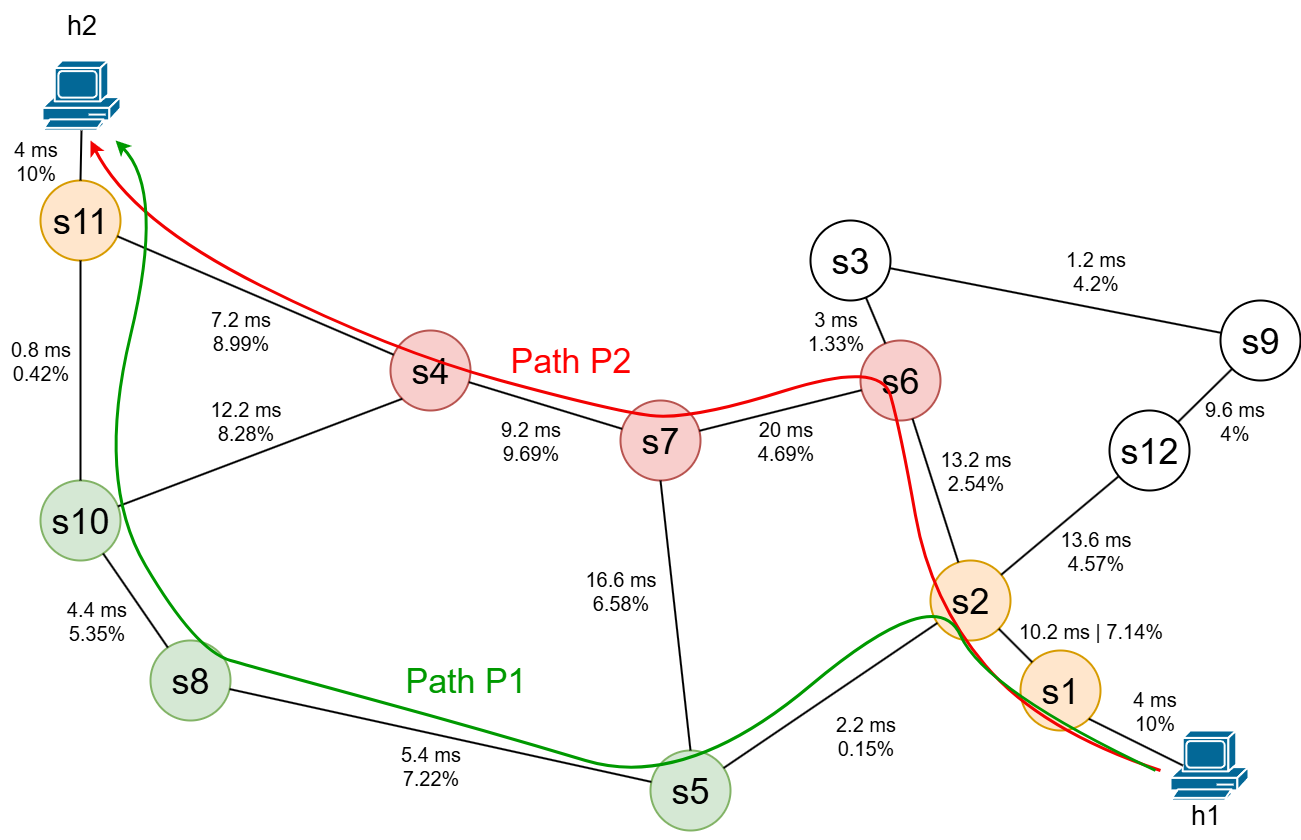}
    \caption{\textcolor{black}{Possible paths for communication between nodes h1 and h2 in the Abilene topology.}}
    \label{fig:path-examples}
\end{figure}

\section{Problem Description} \label{sec:problemDescription}
\color{black}
The performance gains achievable with Transport Assistants critically depend not only on their deployment but also on where they are placed in the network. To illustrate this, we extend the analysis in~\cite{JaimeTAPlacementNOMS2024} using the Abilene topology, shown in Fig.\ref{fig:path-examples}. The network consists of 12 nodes and 15 bidirectional links, each annotated with its propagation delay and its packet loss probability. We consider two end hosts, $h1$ and $h2$, that can communicate through two alternative routes: Path1 (red) and Path~2 (green).

Applying Eq.\ref{eq:expDelivDelay}, we compute the Expected Packet Delivery Delay (EPDD) for each path under different TA placements. The results are summarized in Table~\ref{tab:path1}. These tables report the~EPDD when (i) no TA is deployed and (ii) a TA is placed at~various nodes along the path.

Two key insights emerge from this analysis. First, routing alone already plays a major role in performance. In~our~example, simply switching from Path1 to Path2 without using any~TA reduces the EPDD by more than 50\%. Second, if~a~TA is deployed, the location of the TA along a given path has a~substantial impact. In both paths, deploying a TA consistently improves EPDD compared to scenarios without a TA, with improvements of up to 25\% when the TA is optimally placed (either for Path1 or~Path2 as can be seen in Table~\ref{tab:path1}).

This example highlights the dual challenge faced by operators:(i)~Selecting the most efficient path between endpoints; (ii)~Determining the~best placement for TAs along that path. Both decisions are interdependent. Choosing a suboptimal path or TA location can lead to tens of milliseconds of additional delay even in a small topology such as Abilene. In larger and more complex networks, this trade-off becomes even more critical, underscoring the need for systematic methods that jointly optimize flow routing and TA placement.

To address these challenges, we consider two distinct network scenarios, each reflecting different operational objectives and constraints:

\begin{table}[t]
\centering
\caption{\textcolor{black}{Expected Packet Delivery Delay (EPDD) for different TA placements along Path1 and Path2 (The EPDD accounts for retransmissions)}}
\label{tab:all-paths}

\begin{minipage}{0.48\linewidth}
\centering
\caption*{\textcolor{black}{Path1}}
\begin{tabular}{c|c}
TA placement & EPDD (ms)\\
\hline
S1 &  64.78 \\
S2 &  55.45 \\
S5 &  54.25 \\
S8 &  \textbf{54.01} \\
S10 &  56.37\\
S11 &  62.77\\
No TA &  72.2
\end{tabular}
\label{tab:path1}
\end{minipage}%
\hfill
\begin{minipage}{0.48\linewidth}
\centering
\caption*{\textcolor{black}{Path2}}
\begin{tabular}{c|c}
TA placement & EPDD (ms) \\
\hline
S1 & 141.26 \\
S2 & 131.94 \\
S6 & 122.57 \\
S7 & \textbf{113.92} \\
S4 & 118.33 \\
S11 & 130.74 \\
No TA &  151.82
\end{tabular}
\label{tab:path2}
\end{minipage}

\end{table}

\textbf{$\bullet$ Scenario 1 - Best-effort network:} in this first scenario, the objective is to minimize the average packet delivery delay across all flows in~the~network. This is achieved by carefully selecting the~routes for each flow and deploying a TA when necessary to accelerate retransmissions. In this scenario, no explicit delay guarantees are required, and the focus is solely on reducing latency.

\textbf{$\bullet$ Scenario 2 - QoS-constrained network:} in this scenario, each flow is associated with a delay constraint, defined as an~SLA. Additionally, deploying a TA incurs a cost that~could differ from one node to another. The objective is to maximize the number of flows that satisfy their delay constraints while simultaneously minimizing the deployment costs of TAs and the penalties incurred when SLAs are violated. This scenario models a more realistic environment where both performance and operational cost must be jointly optimized.

\color{black}



\section{Problem Formulation} \label{sec:problemFormulation}
In the following, we model the joint routing and~TA~placement problem to be solved as an ILP. To this end, we first describe how the system is modeled and define the decision variables. We then propose a formulation and an objective function for each of the studied scenarios. 
Table~\ref{tab:notations}~describes all variables used in our formulation. 

\begin{table}[htb]
    \caption{Notation}
    \centering    
    \begin{tabular}{cm{6.5cm}} 
        \hline
        Notation & Description \\
        \hline 
        $\mathcal{N}$ & Nodes of the network \\
        $\mathcal{L}$ & Links of the network \\
        $\bar b_l$ & Bandwidth capacity of link $l$ \\
        $\mathcal{F}$ & Set of flows \\
        $\mathcal{P}_f$ & Set of possible paths to route the flow $f$ \\
        $\mathcal{C}_n$ & Processing capacity of node $n$ \\
        $\mu$ & Upper bound for the number of TAs that should be deployed in the network \\
        $\alpha_n$ & Monetary cost in dollars to process 1~Mbps in~node~$n$ \\
        $s_f, d_f, b_f$ & Source, destination and bandwidth of flow~$f$\\
        $\theta_f$ & Penalty expressed in dollars per ms for not providing the delay required for the flow $f$ \\
        $\bar d_f$ & Upper limit of expected EPDD for flow $f$ \\ 
        $\delta_{f, p}^n$&Expected packet delivery delay for flow $f$ when routed through the path $p$ \\ \hline
        \multicolumn{2}{c}{ILP-specific notation} \\ \hline
        $\beta_{f,p}^n$ & Boolean variable indicating whether the path $p$ used to route the flow $f$ passes through the node $n$ \\
        $\varphi_{f,p}^l$ & Boolean variable indicating whether the link $l$ is used by the possible path $p$ of the flow $f$ \\
        $x_{n}$&Boolean variable indicating whether a TA is instantiated in node $n$ \\
        $y_{f,p}^n$&Boolean variable that indicates that the flow $f$ is routed through the path $p$ and uses the TA instantiated in~node~$n$\\ \hline
        \multicolumn{2}{c}{Heuristic-specific notation} \\ \hline
        $Pair_f$ & Possible path-TA node pairs \((p,n)\) of flow $f$ sorted descending according to the benefit they offer\\
        $B^{\text{rem}}_l$ & Auxiliary variable to account for the available capacity in each link $l$ \\
        $C^{\text{rem}}_l$ & Auxiliary variable to account for the available capacity in each node $n$ \\
        TAs & Nodes that can host the TA function in a run \\
        $S_f$ & Variable that will contain the path-TA pair assigned to flow $f$ as the solution \\
        \hline
    \end{tabular}
    \label{tab:notations}
\end{table}

\textit{\textbf{$\bullet$ System Model:}} We model the network as a bipartite graph $G = (\mathcal{N} , \mathcal{L})$, where $\mathcal{N}$ denotes the set of physical nodes and $\mathcal{L}$ denotes the set of links in the network. Each node $n\in\mathcal{N}$ has a processing capacity limit, $C_n$, that is the maximum amount of bandwidth that could be processed when a TA is placed in node $n\in \mathcal{N}$.  We also define $\bar b_l$ as the bandwidth capacity of~link $l\in\mathcal{L}$.


We define $\mathcal{F}$ as the set of flows. Flow characteristics are captured using the following parameters. Variables $s_f$ and $d_f\in\mathcal{N}$ indicate the source and~destination nodes of the flow $f$, respectively. Variable~$b_f$ is the bandwidth requirement of the flow $f$ expressed in~Mbps. For~each flow $f\in \mathcal{F}$, we define a set of possible paths~$\mathcal{P}_{f}$ through which it could be routed \textcolor{black}{(i.e., possible paths are easily pre-computed using existing algorithms~\cite{cormen2009introduction, yen1971finding,eppstein1998finding})}. The boolean parameter $\beta_{f,p}^n\in \{0,1\}$ indicates whether the path $p \in \mathcal{P}_{f}$ passes through the physical node $n$.
We also define $\varphi_{f,p}^l\in\{0,1\}$ to indicate if the link $l$ is used by path~$p\in \mathcal{P}_{f}$. 
As previously mentioned, we denote by $\delta_{f, p}^n$ the~expected packet delivery delay for flow $f$ when routed through the path $p$ and~assuming the flow is assigned to a TA located in physical node $n$.  This delay is pre-computed for each path and~TA~location based~on~Eq.~\ref{eq:expDelivDelay}.
We also assume the network operator could provide an~upper bound $\mu$ on the number of TAs that should be deployed in the network.  

\textit{\textbf{$\bullet$ Decision variables:}} to solve the problem, we define two decision variables. 
First, we define $x_{n}\in \{0,1\}$ to indicate whether a TA is instantiated in node $n\in\mathcal{N}$. Note that a TA could handle multiple flows. The amount of resources consumed by the TA is proportional to the bandwidth required by the flows it is handling but could not exceed the capacity of the hosting node.  
We also define a boolean variable $y_{f,p}^n\in\{0,1\}$ that takes 1 if the flow $f$ is routed through the~path $p\in \mathcal{P}_{f}$ and uses a TA function deployed in node $n$.

Furthermore, in order to take into consideration the~case where no TA is needed, we added a fictive node $n^*$ to~the~set of~nodes~$\mathcal{N}$ where $\delta_{f, p}^{n^*}$ is the end-to-end delay of~the~path~$p$. This node belongs to all paths ($\beta_{f,p}^{n^*}=1\ \forall\ p\in \mathcal{P}_{f}$), and~has unlimited resources ($C_{n^*}=+\infty$ i.e.,~big value). Hence, when the fictive node $n^*$ is selected by the ILP to~host a~TA for~a~particular flow $f$, it~means that there~is~no~need to~deploy a~TA for~that flow. In this case, the~variable $y_{f,p}^{n^*}$ will capture the best path $p$ to be used to route the flow when~no~TA is needed.


\textit{\textbf{$\bullet$ Objective for Scenario~1 - Joint Routing and packet delivery delay minimization:}}
The objective of this first ILP is to minimize the average expected packet delivery delay over all the flows. Thus, the objective function is defined as follows: 
\begin{align}
    \textbf{Obj 1}: \ \ & \textit{Minimize}\  \frac{1}{|\mathcal{F}|}\ \sum_{f\in\mathcal{F}}\sum_{p\in\mathcal{P}_{f}}\sum_{n\in \mathcal{N}} y_{f,p}^n \ .\  \delta_{f, p}^n \label{eq:obj}\\
    \text{s.t.: }      & \sum_{p\in\mathcal{P}_{f}}\sum_{n \in \mathcal{N}} y_{f,p}^n = 1, \quad\forall f \in\mathcal{F} \label{eq:oneTAperFlow} \\  
    & y_{f,p}^n \leq x_n \cdot \beta_{f,p}^n, \ \forall f \in\mathcal{F}, \forall p \in\mathcal{P}_{f}, \forall n\in\mathcal{N} \label{eq:TAinPath} \\
    & x_n \leq \sum_{f\in\mathcal{F}}\sum_{p\in\mathcal{P}_{f}} y_{f,p}^{n}, \quad \forall n\in\mathcal{N} \label{eq:TAifFlow}\\
     & \sum_{f\in\mathcal{F}}\sum_{p\in\mathcal{P}_{f}} y_{f,p}^n \cdot b_f \leq \mathcal{C}_n, \quad\forall n\in\mathcal{N} \label{eq:NodeCapacity} \\
    & \sum_{f\in\mathcal{F}}\sum_{p\in\mathcal{P}_{f}}\sum_{n \in \mathcal{N}} \varphi_{f,p}^l \cdot b_f \cdot  y_{f,p}^n \leq \bar b_l, \quad \forall l\in\mathcal{L} \label{eq:LinkCapacity}\\
    & \sum_{n\in\mathcal{N}}x_n \leq \mu \label{eq:TAupperbound} 
\end{align}

%
The~first constraint (Eq.~\ref{eq:oneTAperFlow}) guarantees that a flow is handled by one single TA. 
Note that a TA could handle multiple flows but the a flow is handled by at most one TA.
In~addition, Eq.~\ref{eq:TAinPath}~ensures that a flow $f$ could be assigned to a TA placed in node $n$ only if its routing path $p \in \mathcal{P}_{f}$ goes through that node. Moreover, the same equation forces $x_n$ to be equal to 1 if node $n$ is used  as a hosting node for the TA supporting flow $f$.
%
The~next constraint Eq.~\ref{eq:TAifFlow} stipulates that a TA is deployed in node $n$ if a flow $f$ is assigned to that node $n$. 


As previously introduced, the TA consumes bandwidth and processing  resources proportional to the number of flows assigned to~it and their bandwidth requirements. However, the amount of~resources consumed by the TA should not exceed the capacity of the hosting node. This is captured by Eq.~\ref{eq:NodeCapacity}, for which the sum of bandwidth of all the flows using a TA does not exceed the capacity of its hosting node $n$. 
%
Eq. \ref{eq:LinkCapacity} ensures that the bandwidth used by flows using a physical link does not exceed its bandwidth capacity.
The last constraint (Eq.~\ref{eq:TAupperbound}) ensures that the number of TA functions deployed in the whole network is less than an upper bound $\mu$ defined by the network operator.\\


\textit{\textbf{$\bullet$ Objective for Scenario~2 - Joint Routing and TA placement considering SLA and costs:}}
In~the~second scenario, we consider more constrained environments where an SLA is defined stipulating that, for each flow $f \in \mathcal{P}_{f}$, the packet delivery delay should not exceed an upper bound~$\bar d_f$. If this constraint is not satisfied, a penalty is applied for each delay difference with respect to the specified upper bound. Let $\theta_f$ denote the~penalty expressed in dollars per ms, which can be specified by the operator depending on the desired QoS for the flow. 
The total penalty amount for all flows could be then computed as follows:

\begin{equation}\label{DelayPenalty}
\sum_{f\in\mathcal{F}}\sum_{p\in\mathcal{P}_{f}}\sum_{n\in \mathcal{N}} y_{f,p}^n \ .\ \theta_f .\ max(\delta_{f, p}^n - \bar d_f,0)
\end{equation}

In this formulation, we also consider minimizing the deployment costs of TAs. We hence define $\alpha_n$ as the monetary cost in dollars to process 1 Mbps of data at~node $n\in \mathcal{N}$. Naturally, the processing cost varies from one node to another depending on many parameters like the energy costs and operational expenses in the node's location.  Hence, the deployment cost of one TA will be proportional to~$\alpha_n$ as~well as~the~bandwidth of the flows that it handles. Note that for the fictive node~$n^*$, $\alpha_{n^*}$ is equal to zero as placing the TA in $n^*$  means that no TA is deployed.
The total deployment cost of TAs can be computed as follows:
\begin{equation}\label{TADeploymentCosts}
    \sum_{f\in\mathcal{F}}\sum_{p\in\mathcal{P}_{f}}\sum_{n\in \mathcal{N}} y_{f,p}^n \ .\  \alpha_n\ .\ b_f
\end{equation}

Using Eq.~\ref{DelayPenalty} and Eq.~\ref{TADeploymentCosts}, the objective of the ILP for the second scenario is defined as~follows:


\begin{align}
\textbf{Obj 2:}\quad \textit{Minimize} & 
\sum_{f\in\mathcal{F}} \sum_{p\in\mathcal{P}_{f}} \sum_{n\in\mathcal{N}}
y_{f,p}^n \Big[ \alpha_n b_f  \nonumber\\
&\qquad + \theta_f \max\!\big(\delta_{f,p}^n - \bar d_f,\, 0\big) \Big]
\label{eq:Obj2}
\end{align}

This objective is also subject to constraints Eq.~\ref{eq:oneTAperFlow} to~\ref{eq:TAupperbound}. \\

\textcolor{black}{The problem of jointly routing flows and selecting TA placements combines routing decisions with network function placement. However, both problems are computationally intractable in their general forms. In particular, QoS-constrained routing and network function placement have been shown to~be NP-hard~\cite{LuizelliIM2015}. Consequently, in the following section, we~propose a heuristic algorithm to obtain near-optimal solutions in~tractable time.}

\section{Heuristic Algorithm for the joint Routing and~TA placement problem}\label{sec:heuristics}

In this Section, we propose a heuristic framework called \emph{Transport Assistant Fast Selection} (TAFS), which can be adapted to solve either Objective~1 or Objective~2. When focusing specifically on minimizing the average expected packet delivery delay (Objective~1), we refer to the variant as TAFS1. When addressing the objective of minimizing the expected penalty (Objective~2), the corresponding variant is denoted TAFS2. The pseudo-code of the general TAFS heuristic is described in Algorithm~\ref{alg:Heuristic-1}.

The heuristic assigns flows to feasible $(p,n)$ pairs (path and TA-node) while respecting node and link capacities, as well as the upper bound $\mu$ on the number of TAs. \textcolor{black}{The main idea is to process flows in order of their size ($b_f$), and for each flow, process pair $(p,n)$ in order of their potential benefit}: for~TAFS1, the benefit corresponds to the expected reduction in packet delivery delay from using a TA, whereas~for~TAFS2, it corresponds to the expected reduction in penalty. Flows are then assigned sequentially to the best available option according to the chosen objective.

More formally, the proposed algorithm operates as follows. For each flow \( f \in \mathcal{F} \), every candidate path \( p \in \mathcal{P}_f \), and each potential TA node \( n \in \mathcal{N} \) located along \( p \) (including the fictive node \( n^* \)), the expected packet delivery delay \( \delta_{f,p}^n \) is assumed to be known. 
We define the \emph{benefit} of selecting a particular path \( p \) and~placing the TA at node \( n \) as the improvement gained relative to the best available solution without deploying a~TA~on~that~path. 
For TAFS1, it could be computed as follows in order to prioritize reducing delays: 
\begin{equation}
  \text{benefit}_{f,p,n}^{\text{TAFS1}}  = 
  \delta_{f,p}^{n^*} - \delta_{f,p}^n ,
     \label{eq:benefit-tafs1}
\end{equation}
where $\delta_{f,p}^{n^*}$ is the delay when no TA is used and when path $p$ is considered to route flow~$f$.  

However,  TAFS2 prioritizes reducing penalties, and hence, it is computed as follows: 
\begin{align}
\text{benefit}_{f,p,n}^{\text{TAFS2}} 
   &= \max\Big(0, \; 
      \theta_f \cdot \max(0, \delta_{f,p}^{n^*} - \bar d_f) \nonumber \\
   &\quad - \theta_f \cdot \max(0, \delta_{f,p}^n - \bar d_f)\Big),
   \label{eq:benefit-tafs2}
\end{align}
%

\algrenewcommand\algorithmicindent{1.4em} 

\begin{algorithm}[H]
\caption{Pseudo-code of TAFS.}\label{alg:Heuristic-1}
\begin{algorithmic}[1]
\State Initialize residual node capacities $C^{\text{rem}}_n \gets C_n$, 
       residual link capacities $B^{\text{rem}}_l \gets \bar b_l$, pair \((p,n)\) assign to each flow $S_f \gets \varnothing$ for all $f \in \mathcal{F}$
\State Set TAs $\gets orderedTAs[:\text{number of TAs}]$
\State Order flows in decreasing order of $b_f$
\For{each flow $f$ in that order}
   \For{each $p,n \in Pairs_f$}
      \If{($n \in $ TAs \textbf{or} $n = n^*$) \textbf{and} for all $l$ in the path $p$ : $B^{\text{rem}}_l \ge b_f$}
        \If{$n \in TAs$ and $C^{\text{rem}}_n < b_f$} \textbf{continue} 
        \Else \text{ } $C^{\text{rem}}_{n} \gets C^{\text{rem}}_{n} - b_f$ \EndIf
        \State Update capacities $B^{\text{rem}}_l$ of all~links~$l$ in $p$
        \State $S_f \gets [p, n]$
        \State \textbf{break}
      \EndIf
   \EndFor
    \If{$S_f = \varnothing$}
      \State Assign flow $f$ as rejected
   \EndIf
\EndFor
\State \Return TAs and $S_{f\in\mathcal{F}}$
\end{algorithmic}
\end{algorithm}

The proposed algorithm (Algorithm~\ref{alg:Heuristic-1}) operates in several main stages.  
First, it initializes the residual node capacities \(C^{\text{rem}}_n\), the residual link capacities \(B^{\text{rem}}_l\) to their total capacities, and \textcolor{black}{the solution list that will contains the \((p,n)\) pair assigned to each flow} (Line~1)
\textcolor{black}{Prior to the execution of the algorithm, for each flow, the benefits of using every candidate path \( p \in \mathcal{P}_f \), and each potential TA node \( n \in \mathcal{N} \) located along \( p \) (including the fictitious node \( n^* \)) ($\text{benefit}_{f,p,n}^{\text{TAFS1}}, \forall f \in \mathcal{F}, p\in\mathcal{P_f},n\in\mathcal{N}\cup n^*$), must be precomputed ($Pairs_f$).}
Here, the benefit depends on the chosen variant of TAFS (TAFS1 or TAFS2), as defined in~Eq.~\ref{eq:benefit-tafs1}~and~\ref{eq:benefit-tafs2}.  
After computing the benefits, \textcolor{black}{for each flow, the possible pairs \((p,n)\) are sorted in descending order of benefit, which enables a fast search among all possible solutions: once a valid path and TA node are selected, none of the subsequent options can improve the benefit of that flow.} \textcolor{black}{At this stage, for each node \(n \in \mathcal{N}\), the average benefit is computed over all candidate paths of each flow that use it as a TA (Lines~2). The goal is to rank the nodes in descending order of average benefit ($orderedTAs$), so that TAs can be deployed on the nodes that, on average, provide the highest benefit, up to the number of TAs specified by the administrator.}

For each flow (\textcolor{black}{sorted in descending order of size}) (Lines~4–18), the algorithm searches over all feasible pairs \((p,n)\) pairs (Line~5). 
Feasibility is checked against residual link capacities (Line~6) and residual node capacities if the TA is deployed (Line~7). 
\textcolor{black}{When the algorithm finds a pair \((p,n)\) that satisfies these constraints, it updates the node capacities (Line~8), updates the residual link capacities along the chosen path (Line~10), assigns the flow to that \((p,n)\) path/TA placement pair (Line~11), and stops the search for that flow (Line~12), since the pairs are ordered by the benefit they provide and this one is the best possible for the current network state.}
If no feasible \((p,n)\) pair exists, the algorithm assigns the flow as rejected (Line~16).  

Finally, \textcolor{black}{the algorithm returns the nodes where the TAs will be deployed and the \((p,n)\) pair assigned to each flow.}
This design ensures that TAFS respects capacity constraints and the TA deployment limit $\mu$, while prioritizing \textcolor{black}{pairs \((p,n)\)} based on their expected benefit according to the chosen objective.
The computational complexity of this algorithm is \textcolor{black}{$O\!\left(\sum_{f \in \mathcal{F}} |\mathcal{Pair}_f| \cdot |\mathcal{N}| \cdot |\text{len(}p_f\text{)}^2| \right)$, assuming the set of pairs \((p,n)\) for each flow is precomputed.}





\section{Performance Evaluation}\label{sec:results}

In order to evaluate the performance of the proposed solutions, we run a set of experiments using a variety of scenarios considering different network topologies. 
We first describe the simulation setup and then present the performance evaluation results of~the~ILP solutions using the ILP Solver Gurobipy~\cite{Gurobipy} and then compare them with the proposed heuristic~TAFS. 

\subsection{Simulation setup}


In the conducted simulations, we consider three network topologies of different sizes: Abilene (12~nodes, 30~links), Geant (22~nodes, 72~links), and Germany (50~nodes, 176~links), which are retrieved from~\cite{SNDlib2010}. 
Each node in the network has computing and networking resources that could be used to host a TA. Processing capacities of the nodes are expressed in Mbps. These processing capacities are randomly generated between 200 and 350~Mbps. Furthermore, the processing cost could vary from one node to another depending on the location of the node as the cost of energy and~operational expenditures differs from one node to another. In~our~experiments, the processing cost for each node was randomly generated between $7\cdot 10^{-5}$ and~$11\cdot 10^{-5}$~dollars~per~Mbps. This cost range is the closest to the one proposed for AWS t4g.micro virtual machines\footnote{\url{https://aws.amazon.com/ec2/pricing/on-demand/}}.





\begin{figure*}[!h]
    \begin{center}
    \subfloat[Average EPDD vs.~number of nodes with TAs (Abilene topology).]{\includegraphics[width=0.3\textwidth]{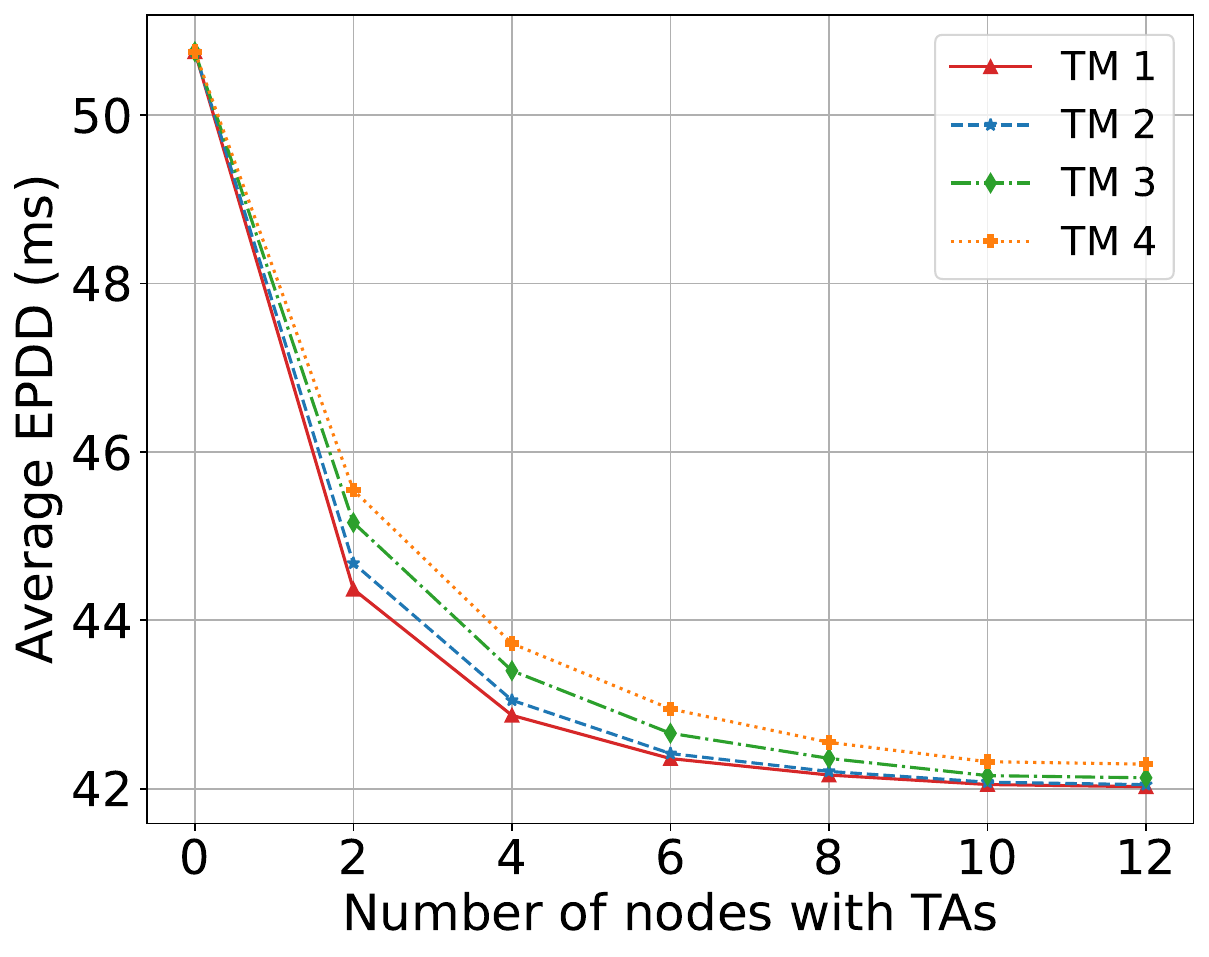}\label{fig:PacketDeliveryDelay-obj1}}
    \hspace{4pt}
    \subfloat[EPDD improvement vs.~number of nodes with~TAs (Abilene topology).]{\includegraphics[width=0.3\textwidth]{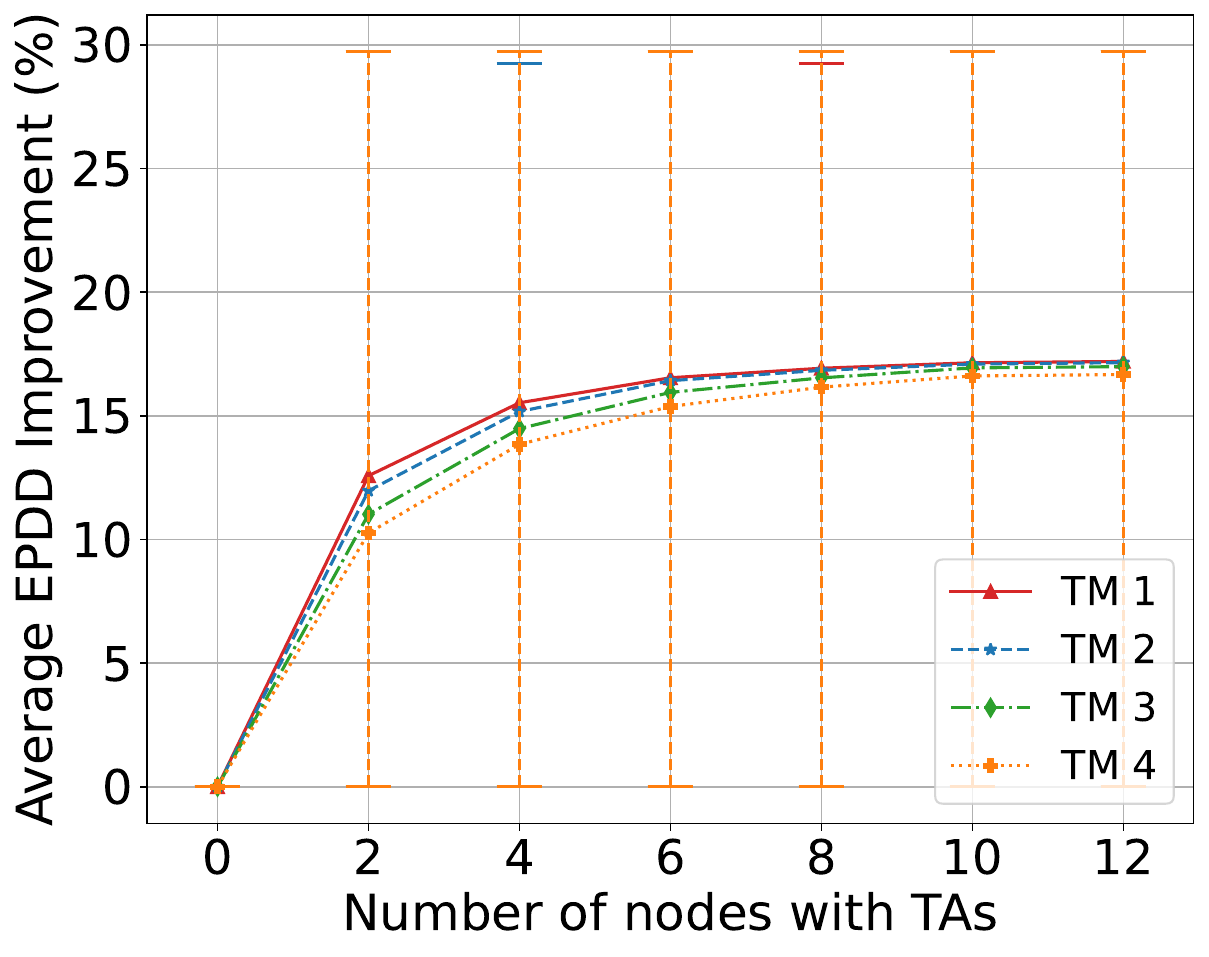}\label{fig:PacketDeliveryDelayImprovement-obj1b}}
    \hspace{4pt}
    \subfloat[CDF of expected packet delivery delay improvement (Abilene topology, TM4).]{\includegraphics[width=0.3\textwidth]{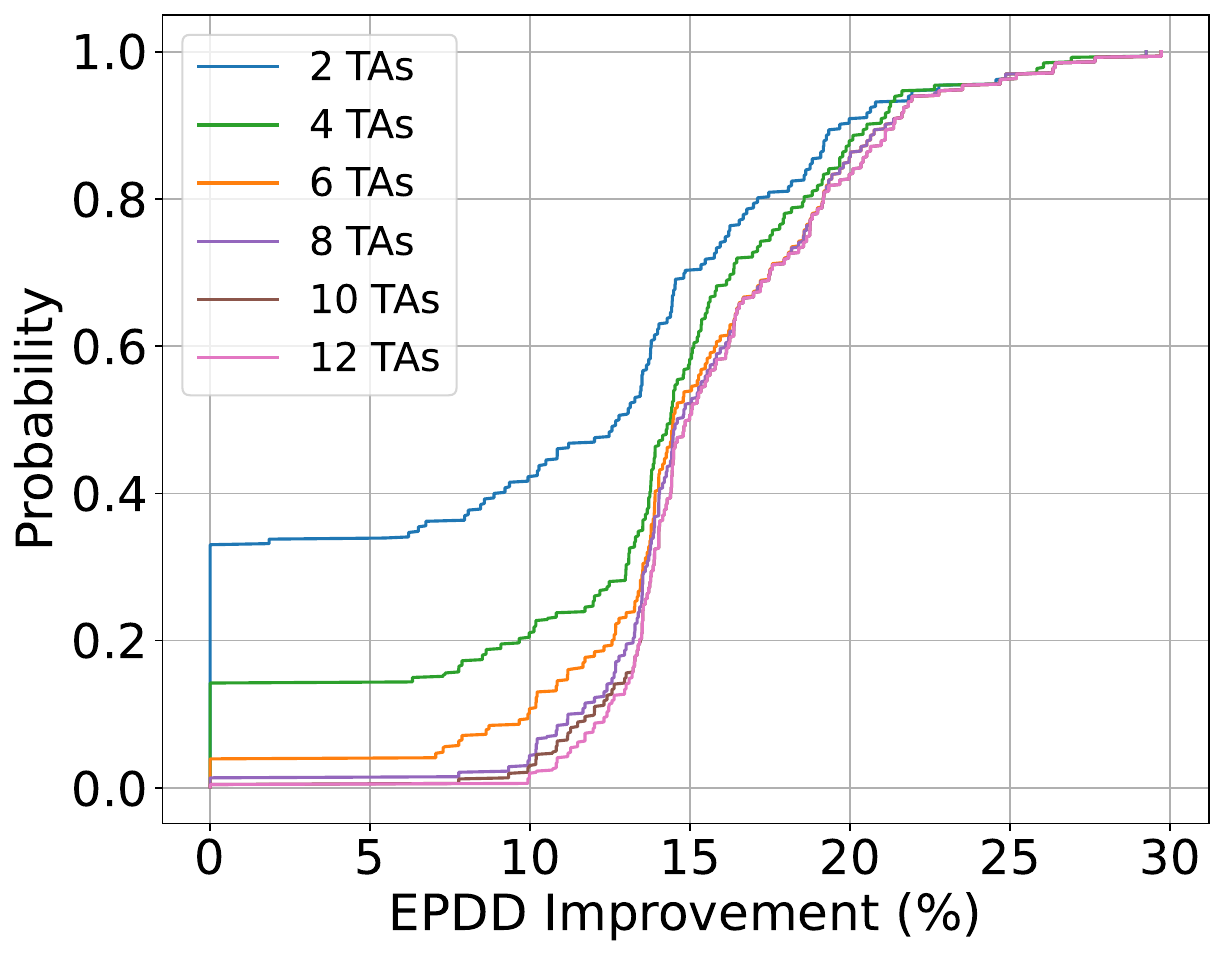}\label{fig:CDF-obj1}}
    \hspace{4pt}
    \subfloat[Total costs  vs.~traffic matrices\\(Abilene topology and 12 nodes with TAs).\\]{\includegraphics[width=0.3\textwidth]{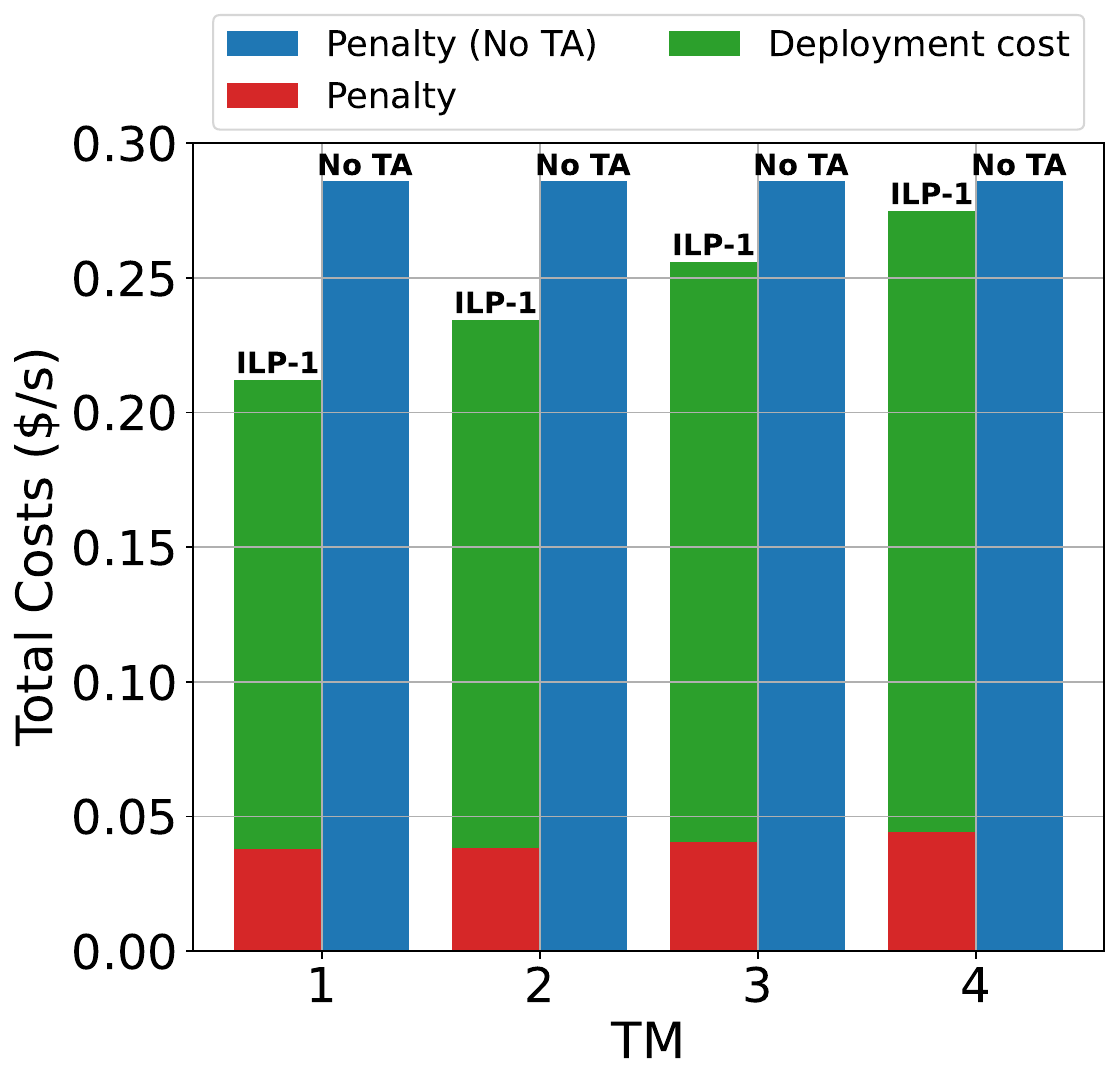}\label{fig:Costs-obj1}}
    \hspace{4pt}
    \subfloat[Total costs vs.~~number of nodes with~~TAs (Abilene topology).\\]{\includegraphics[width=0.3\textwidth]{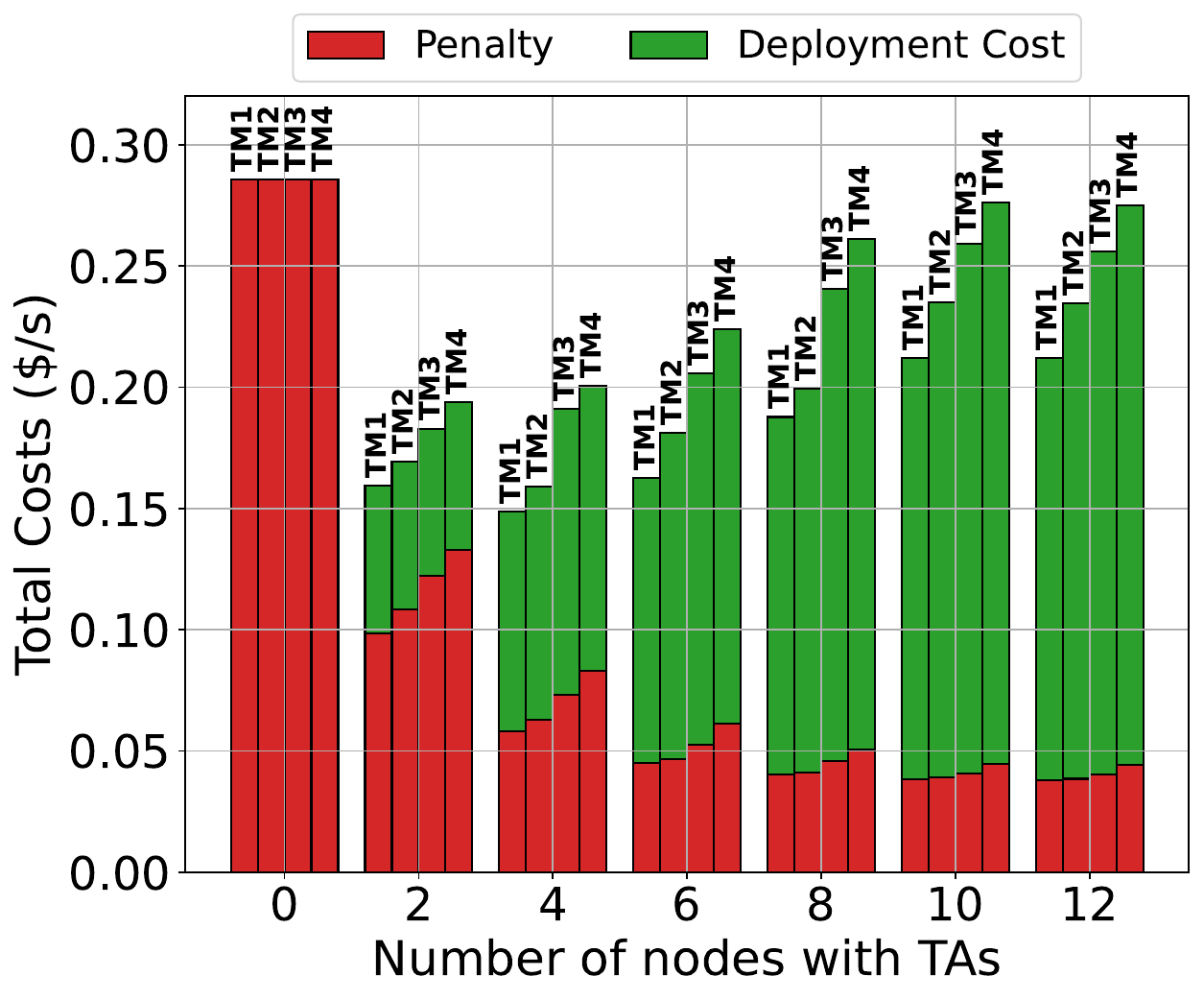}\label{fig:Costs-nodes-TM-obj1}}
    \hspace{4pt}
    \subfloat[EPDD improvement  vs.~~percentage of nodes with TAs for the three network topologies.]{\includegraphics[width=0.3\textwidth]{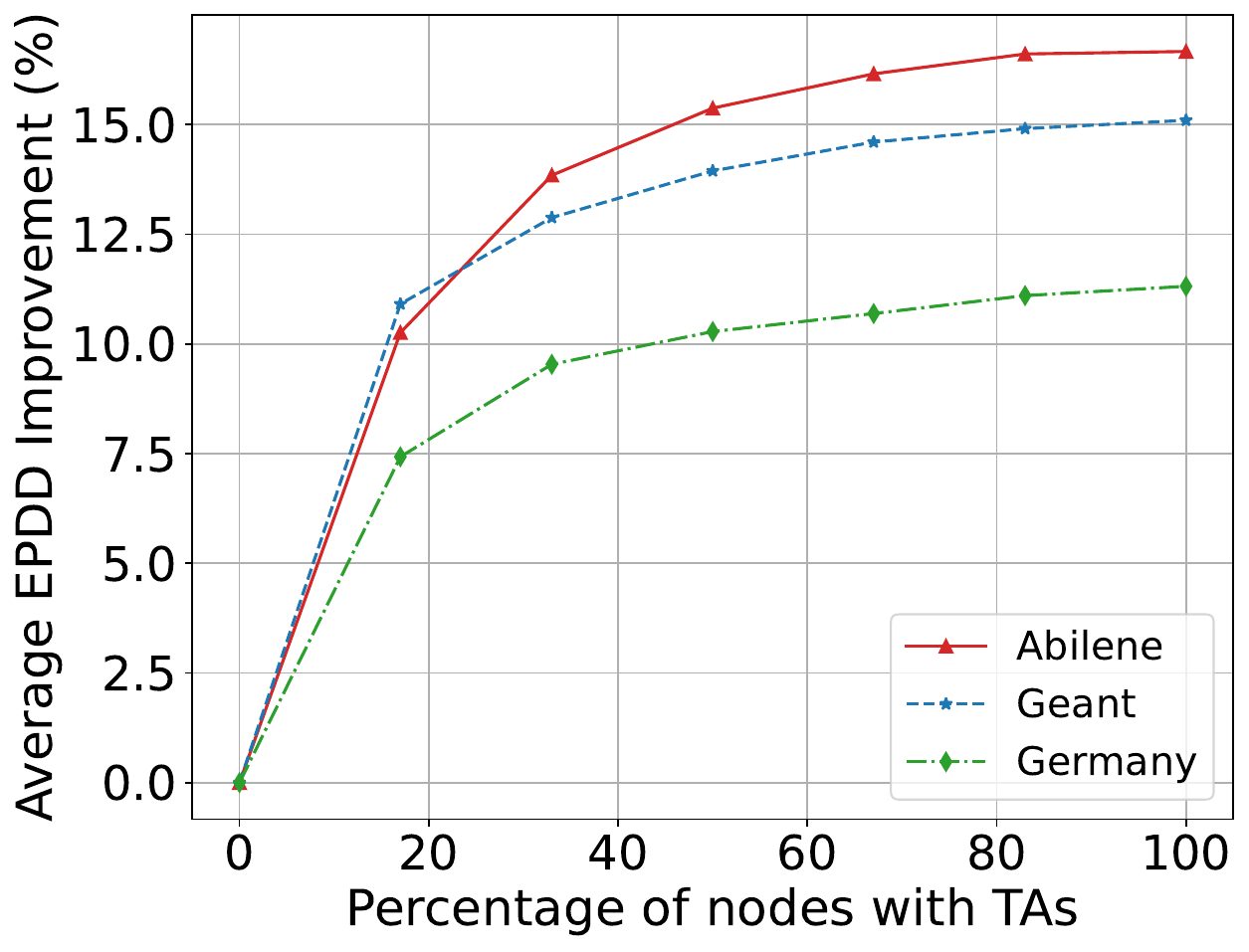}\label{fig:PacketDeliveryDelayImprovement_multitopo-Percents-obj1}}
    \hspace{4pt}
    \caption{Results for Objective 1 (ILP1)}
    \label{fig:Objective1Results}
    \end{center}
\end{figure*}

\color{black}
Furthermore, we assume there are five different flows between each pair of nodes, each with its own SLA that specifies the bandwidth and delay requirements depending on the needs of its associated application. Note that a~flow in~this context could be a single TCP flow or a bundle of TCP flows that have the~same performance requirement, share the same routing path and could be eventually handled by~the~same~TA. 

We consider different Traffic Matrices (TMs) in our evaluation. A~TM specifies the traffic demand between each pair of nodes in the network and thus provides an estimate of the bandwidth requirements for flows. To build these matrices, we randomly set the estimated bandwidth requirement $b_f$ for each flow~$f$. For the Abilene network, we rely on four different TMs to study how varying traffic configurations may influence the results. For the Geant and Germany networks, however, we use a single TM, since our main objective for such large topologies is to evaluate the scalability of the proposed solutions.


Furthermore, the delay requirement $\bar d_f$ for a flow $f$ is randomly set between 90\% and 110\% of~the~shortest path delay between the source and destination. This reflects a realistic case where the flow requirement may be satisfied or not depending on the delay of the available paths between the source and destination. When the delay requirement is not satisfied, a~penalty proportional to the delay difference between the target and the current one is applied as~expressed in~Eq~\ref{DelayPenalty}. The penalty is fixed to $5\cdot 10^{-5}$~dollars for each ms of delay difference.\\
\color{black}

\subsection{Results for the ILP1}

In the first set of experiments, we focus on evaluating the~performance of the ILP1 to achieve Objective 1 and show how the joint routing and TA deployment and placement could minimize the packet delivery delay. Note that penalties and deployment costs are not considered in ILP1. 

Figure~\ref{fig:Objective1Results} summarizes the results. 
For instance, Fig. \ref{fig:PacketDeliveryDelay-obj1} and \ref{fig:PacketDeliveryDelayImprovement-obj1b} show respectively the average EPDD for the Abilene network for different number of TAs (controlled by the parameter $\mu$ defined in ILP1 as the upper bound for the number of TAs) and the percentage of delay improvement compared to the case where no TA is deployed. The figures clearly show that the average EPDD decreases when more TAs are deployed, starting from an improvement of $12.57\%$ when 2~TAs are deployed to reach 16.4\% starting from 8~TAs deployed.
Furthermore, Figure~\ref{fig:PacketDeliveryDelayImprovement-obj1b} also shows the lowest and highest EPDD improvements for all paths in the network (see vertical lines representing the 95\% confidence intervals). It~can be~seen that for some paths the improvement is able to approach~30\%. 
We also note that the improvement is consistent for all the~considered traffic matrices (TM1 to TM4). 

Figure~\ref{fig:CDF-obj1} shows the Cumulative Distribution Function (CDF) of the EPDD improvement for the flows considering TM4 in Abilene network. 
For instance, when 2 TA are deployed, 35\% of the flows do not benefit from any improvement, 50\% of the flows benefit from 10\% to 20\% delay improvement, and finally 15\% of the flow benefit from 20\% to 30\% delay improvement. The more deployed TAs, the more flows that benefit from the EPDD improvement. For instance, if we look at Figure~\ref{fig:CDF-obj1}, in which 12 TAs are deployed, almost all flows experience a delay reduction of 10\%, with over 85\% showing an improvement between 10\% and 20\%. This~represents about 30\% more flows than the case in which only 2 TAs are deployed.

Figure ~\ref{fig:Costs-nodes-TM-obj1} shows how deployment costs and penalties evolve as a function of the number of TAs (i.e.,~network nodes hosting the TAs). When TAs are not used, the operational costs are primarily penalties resulting from SLA violations; However, we can see in the figure that, as more TAs are deployed, the deployment costs grow but the penalties are reduced as the EPDD is decreasing. Naturally, the total operational cost increases when more TA are deployed. It is up then to the operator to decide on how many TAs could be deployed depending whether the objective is to minimize the EPDD and the penalty (according to the figure, 12~TAs is the best option in this case)  or~to~minimize the total operation costs (2 TAs would be sufficient in this case).

Although ILP1 does not aim to minimize the total costs, we depict in Figure~\ref{fig:Costs-obj1} the total costs (Eq.~\ref{eq:Obj2}) i) without considering TAs, and ii) when TAs are used (assuming the optimal number of TAs and the placement found with~ILP1). We can first note that, without TAs, the costs are primarily due to the penalties incurred from not meeting the SLA requirements. When TAs are used, the total cost includes both the deployment cost and the penalties. We can see in the figure that the cost savings thanks to TAs range from 2\% when the network is highly loaded (TM4), to 27\% when the traffic load is lower (TM1).

Finally, to evaluate ILP1 for large networks, we run our experiments for bigger topologies, namely Geant and  Germany. Figure~\ref{fig:PacketDeliveryDelayImprovement_multitopo-Percents-obj1} reports the average EPDD improvement for the three studied network topologies, Abilene, Geant and~Germany, with respect to the number of deployed TAs. 
We can see that, for all of them, the benefit of deploying TAs while routing flows significantly results in an improvement between 10\% to 17\%.

\begin{figure*}[!h]
    \begin{center}
    \subfloat[Total costs for different penalty values with and without TAs (Abilene Network, TM4).]{\includegraphics[width=0.22\textwidth]{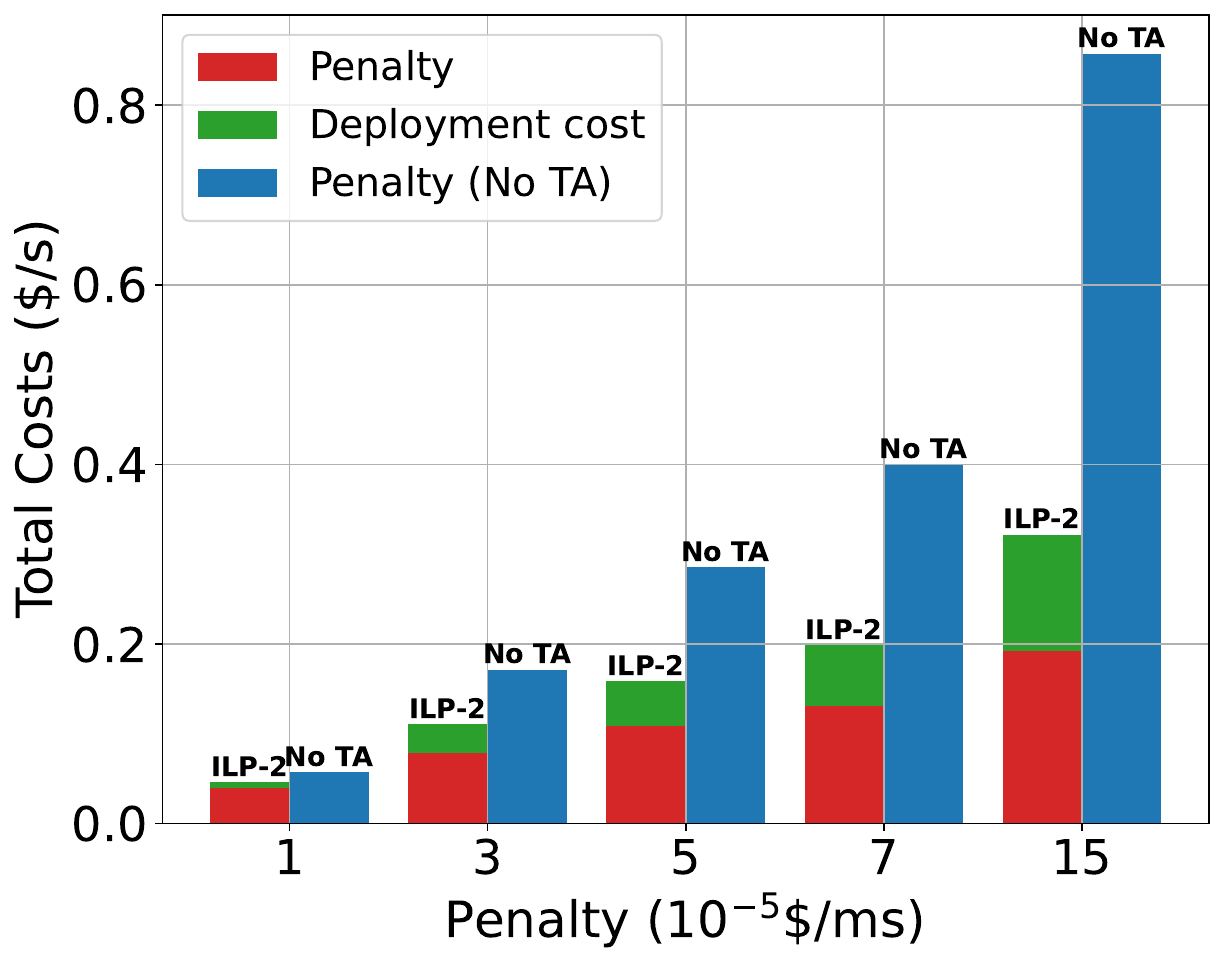}\label{fig:Costs-for-4TM-obj2}}
    \hspace{4pt}
    \subfloat[Total costs with and without TAs for different traffic loads (Abilene Network).]{\includegraphics[width=0.22\textwidth]{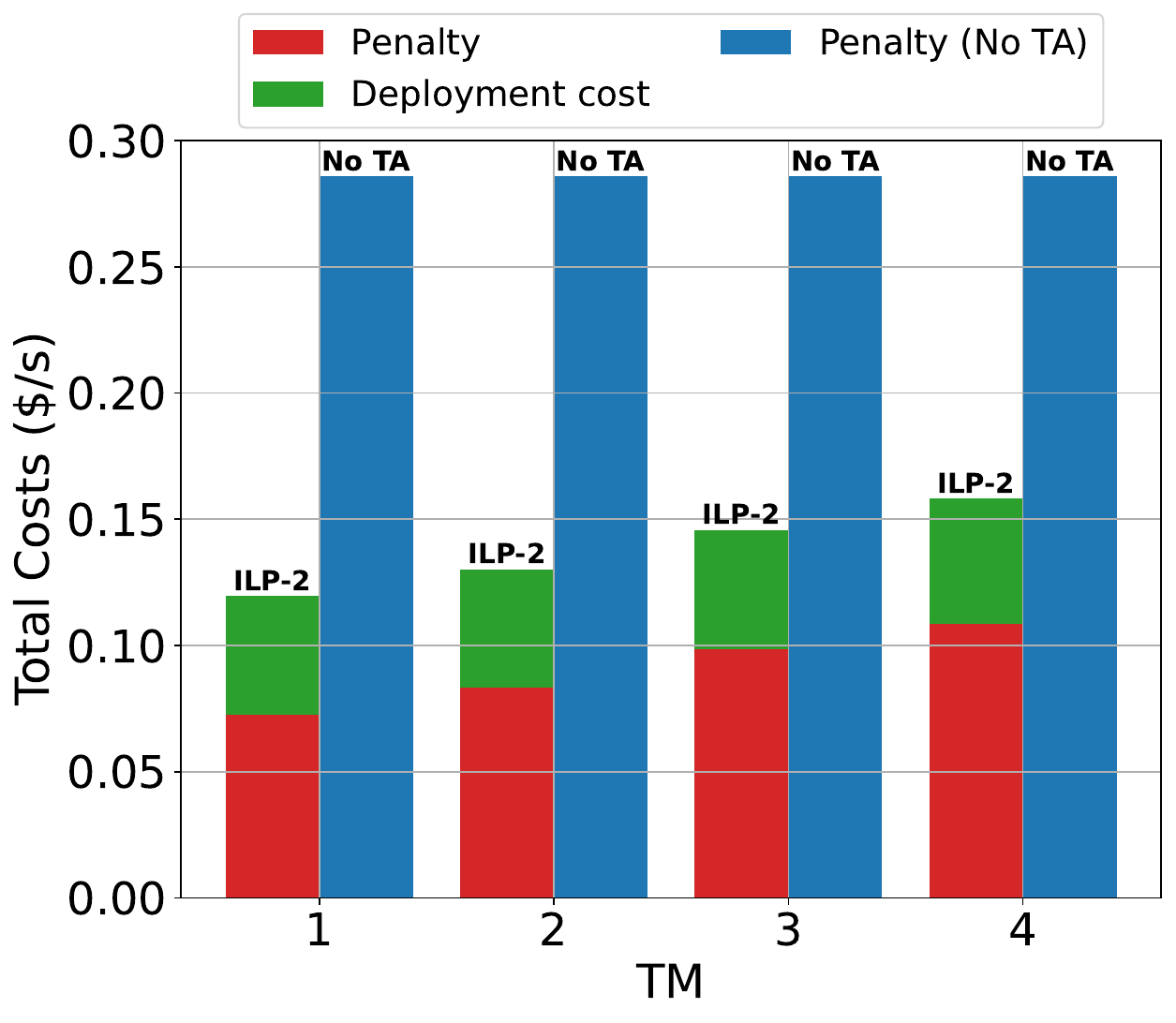}\label{fig:Cost_improvement-obj1b}}
    \hspace{4pt}
    \subfloat[Average EPDD with and without TAs for different traffic loads (Abilene Network).]{\includegraphics[width=0.22\textwidth]{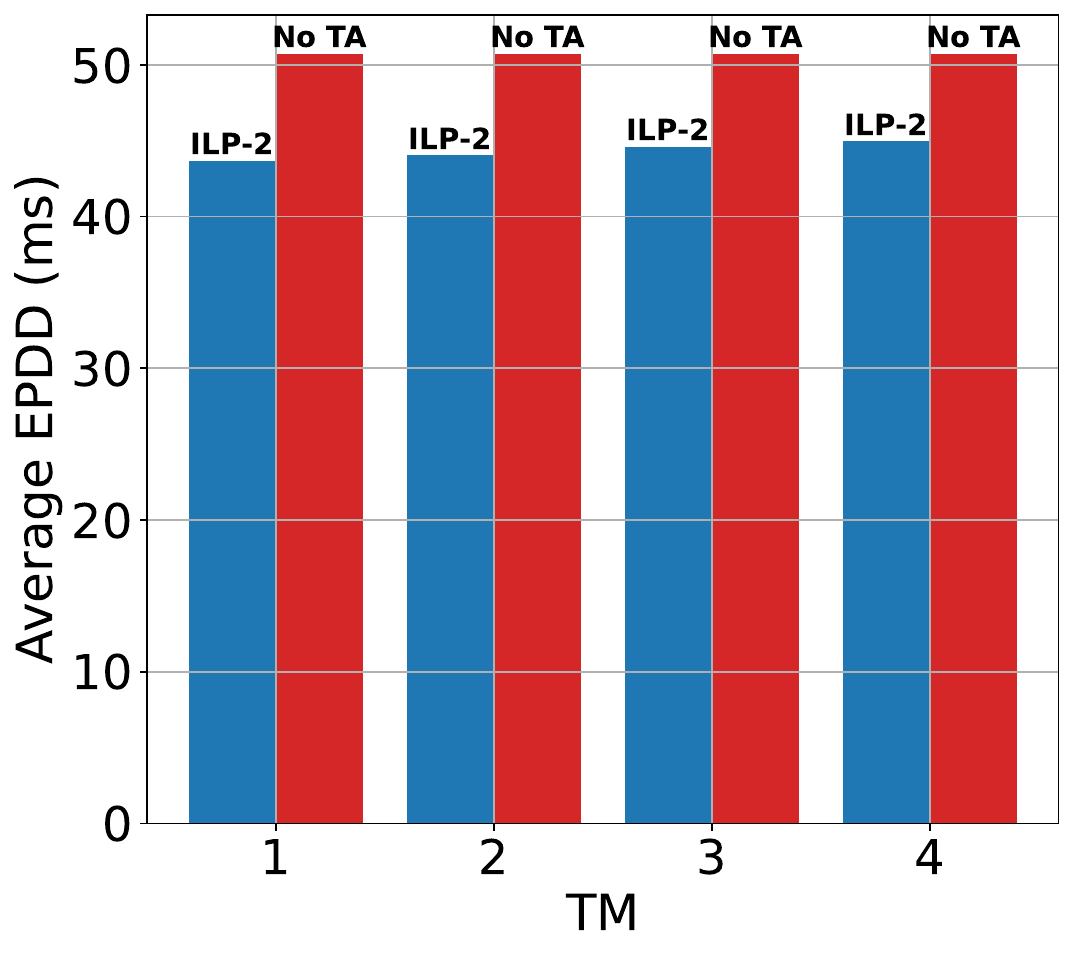}\label{fig:PacketDeliveryDelay-obj2}}
    \hspace{4pt}
    \subfloat[Total costs with and without~TAs for~the~studied topologies (Penalty:~$ 5*10^{-5}$\$/ms)]{\includegraphics[width=0.22\textwidth]{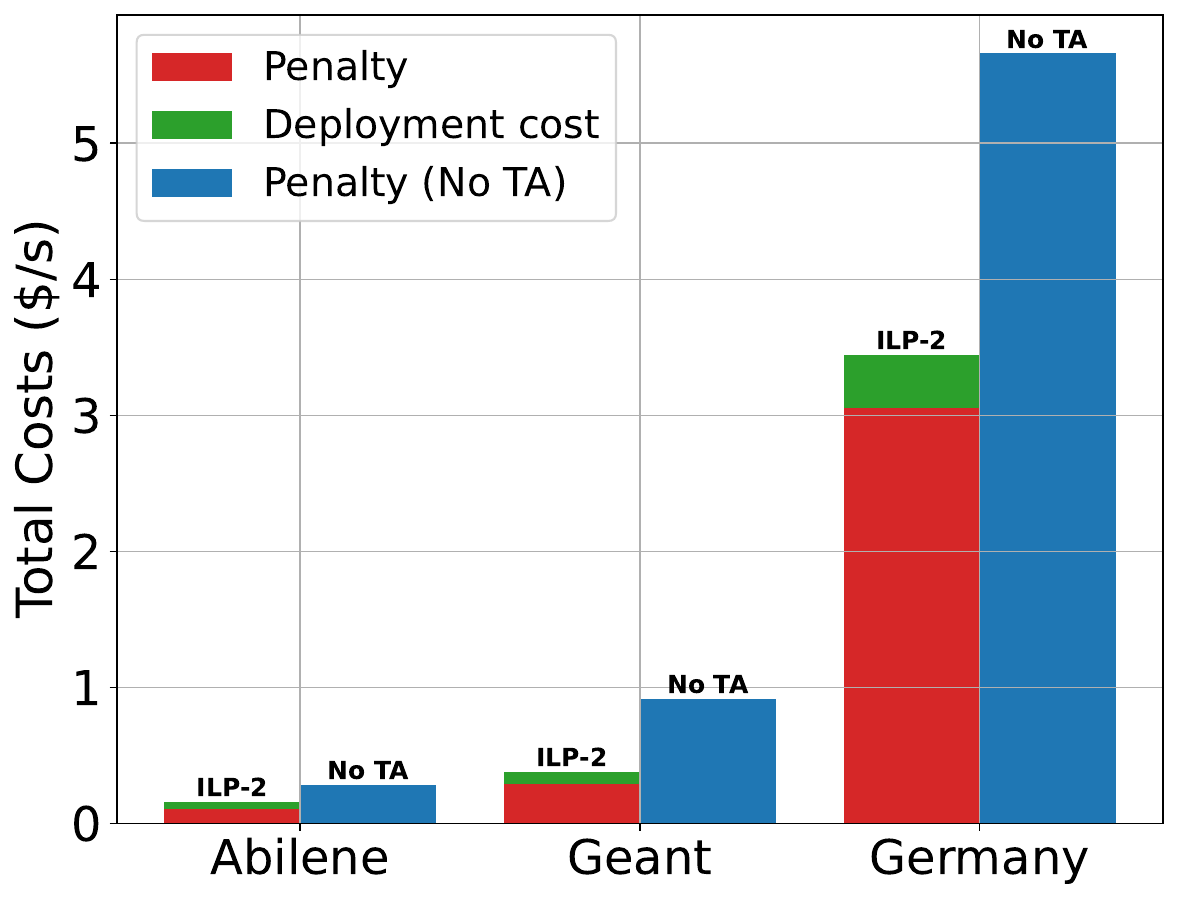}\label{fig:Costs-topologies-obj2}}
    \caption{Results for Objective 2 (ILP2).}
    \label{fig:Objective2Results}
    \end{center}
\end{figure*}

\begin{figure*}[tbh]
    \begin{center}
    \subfloat[Abilene.]{\includegraphics[width=0.3\textwidth]{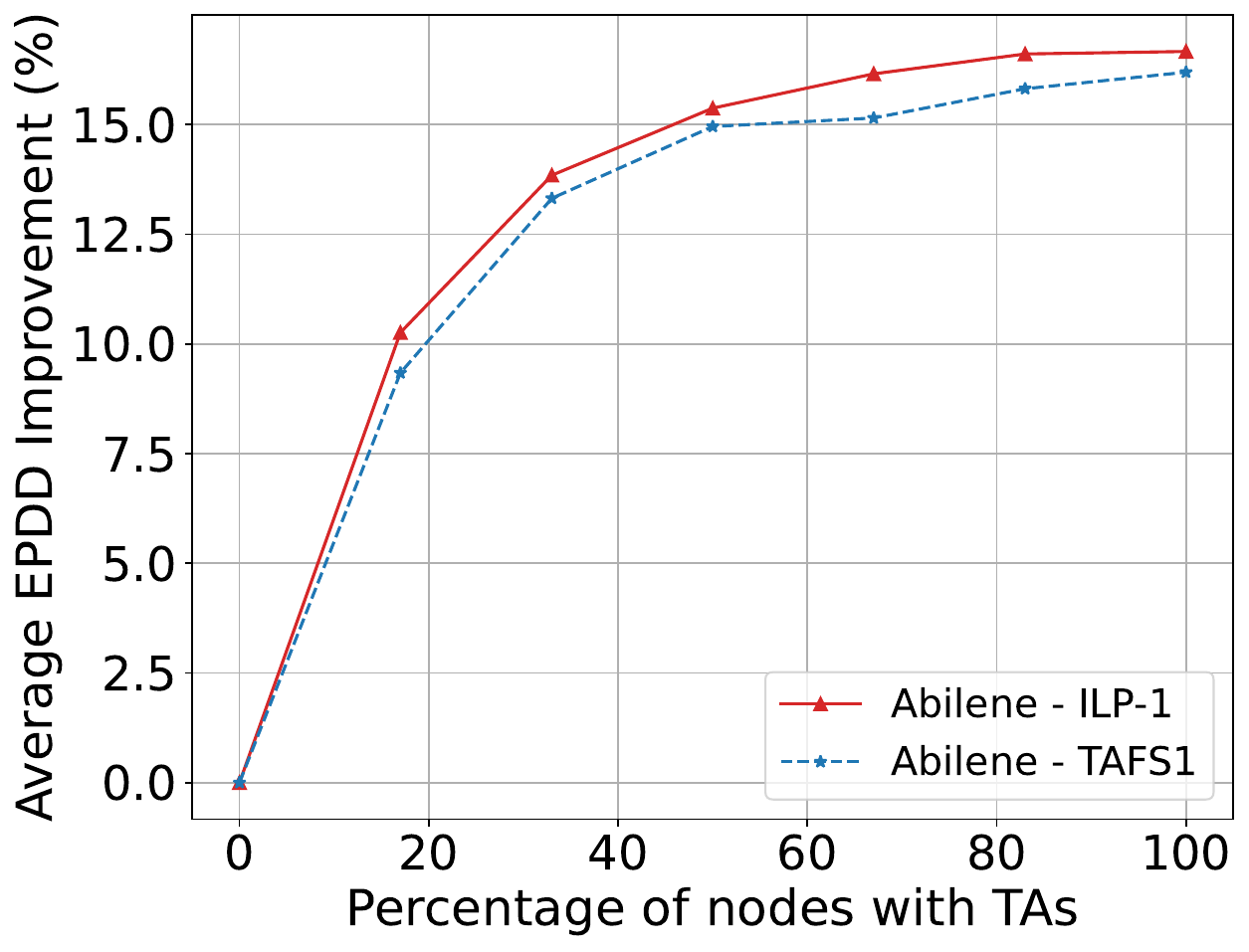}\label{fig:PacketDeliveryDelayImprovement_opt_vs_h-Abilene-obj1}}
    \hspace{4pt}
    \subfloat[Geant.]{\includegraphics[width=0.3\textwidth]{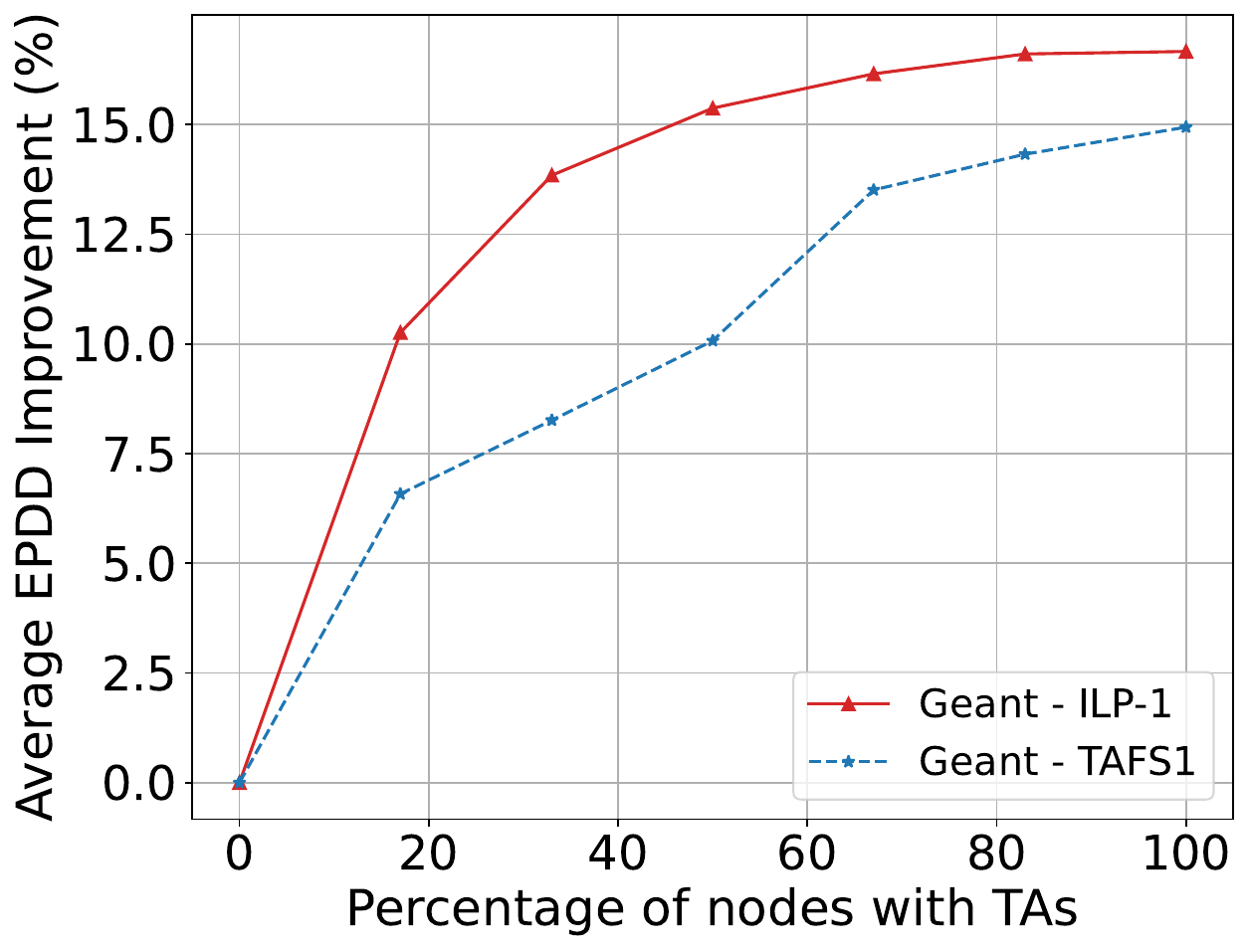}\label{fig:PacketDeliveryDelayImprovement_opt_vs_h-Geant-obj1}}
    \hspace{4pt}
    \subfloat[Germany.]{\includegraphics[width=0.3\textwidth]{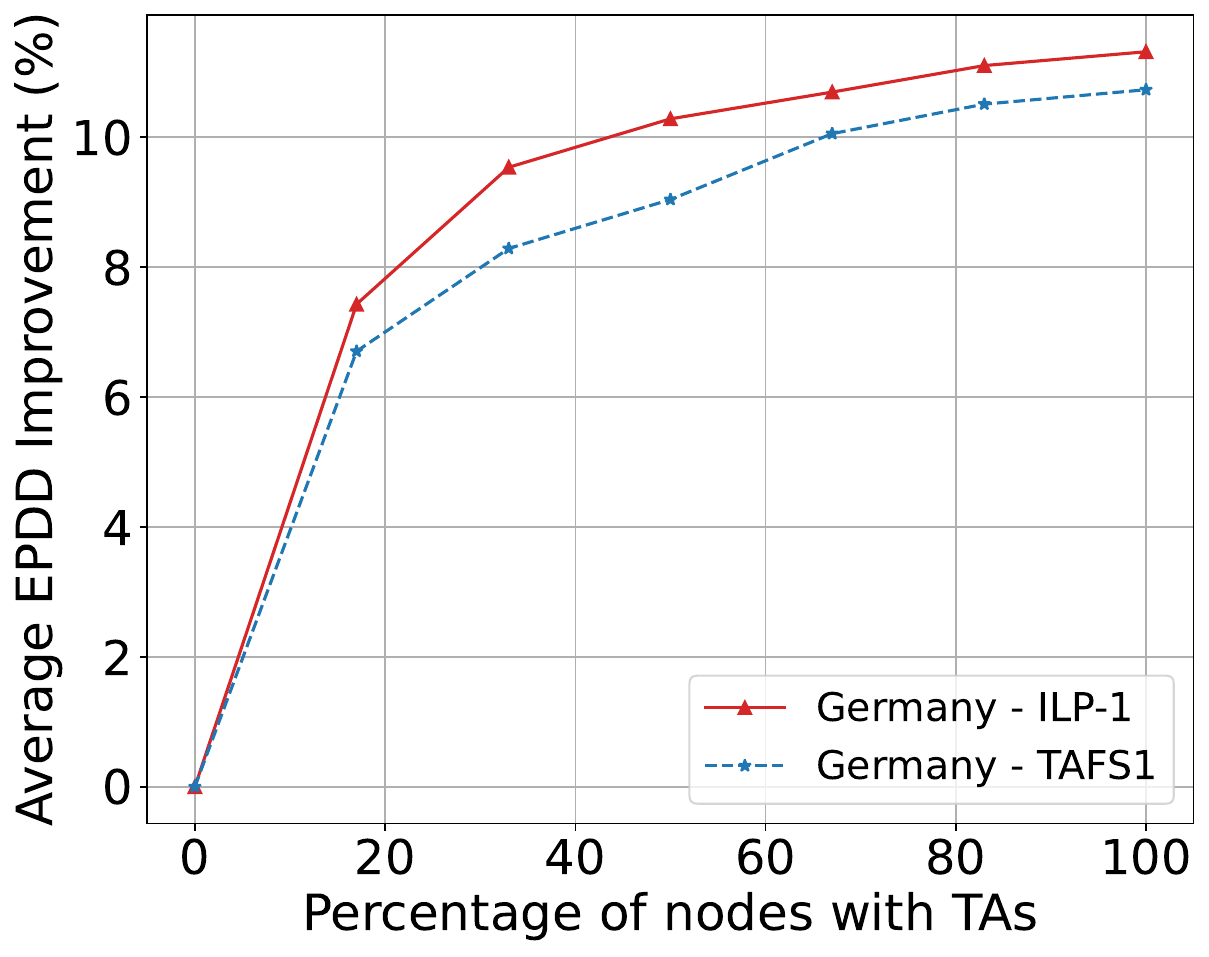}\label{fig:PacketDeliveryDelayImprovement_opt_vs_h-Germany-obj1}}
    \hspace{4pt}
    \caption{Average EPDD improvement vs.~percentage of nodes with TAs for ILP1 and~TAFS1.}
    \label{fig:overall-opt-vs-heuristic-1}
    \end{center}
\end{figure*}

\subsection{Results for the ILP2}
We move now our attention to Objective 2 focusing on the joint routing and TA placement considering a delay-based SLA and its associated penalty as well as the TAs deployment costs. Figure~\ref{fig:Costs-for-4TM-obj2} shows the impact on total costs as a function of the variation of the penalty per ms of delay. We note that the use of TAs significantly reduces the total costs as a consequence of a small increase in the cost due to the deployment of TAs. As we can see, total costs exponentially increase as the penalty associated with SLA non-compliance increases. Hence, the savings when using TAs become much higher and range from 18.22\% to~60.98\% in the best case scenario.

Indeed, Figure~\ref{fig:Cost_improvement-obj1b} shows how costs differ depending on the load of the traffic matrix. It can be seen that the peak TM (TM4) increases the costs up to 0.04\$ per second compared to the least loaded one (TM1).

One step further, Figure~\ref{fig:PacketDeliveryDelay-obj2} shows how ILP2 influences the EPDD and how the resulting values vary depending on the traffic load, improving by 13.92\%, 13.24\%, 12.08\% and 11.42\% the average EPDD compared to not using TAs, for each of the TMs, respectively.

Figure~\ref{fig:Costs-topologies-obj2} shows a comparison of the total costs with TAs (where the optimal number and placement are obtained using ILP2) and without TAs. Clearly, using the set of TAs managed by ILP2 ensures a high reduction in the total costs of 44.67\%, 58.12\%, and 39.14\% for Abilene, Geant, and Germany networks, respectively.
For instance, as shown in the figure, the~use~of~TAs in Germany Network could produce savings of approximately \$2.22 per second which translates into around $70$M\$ of savings per year.



\subsection{Results for the TAFS Algorithm}

Finally, we provide a comparison between our proposed heuristic, TAFS, and the ILPs. \textcolor{black}{
Figures~\ref{fig:PacketDeliveryDelayImprovement_opt_vs_h-Abilene-obj1}, \ref{fig:PacketDeliveryDelayImprovement_opt_vs_h-Geant-obj1}, and~\ref{fig:PacketDeliveryDelayImprovement_opt_vs_h-Germany-obj1} show how TAFS1 (objective 1) achieves near-optimal results, with an optimality gap of 0.69\% for Abilene network, 3.54\% for Geant and 0.84\% for Germany, respectively}. However, the advantage of TAFS is that~it~requires much shorter computation times, a reduction of 902 times for Abilene, 2,256 for Geant, and 3,719 for Germany, as shown in Table~\ref{tab:execution_times}.

\begin{figure*}[h]
    \begin{center}
    \subfloat[Total costs for different penalty values with and without TAs (Abilene Network, TM4)]{\includegraphics[width=0.30\textwidth]{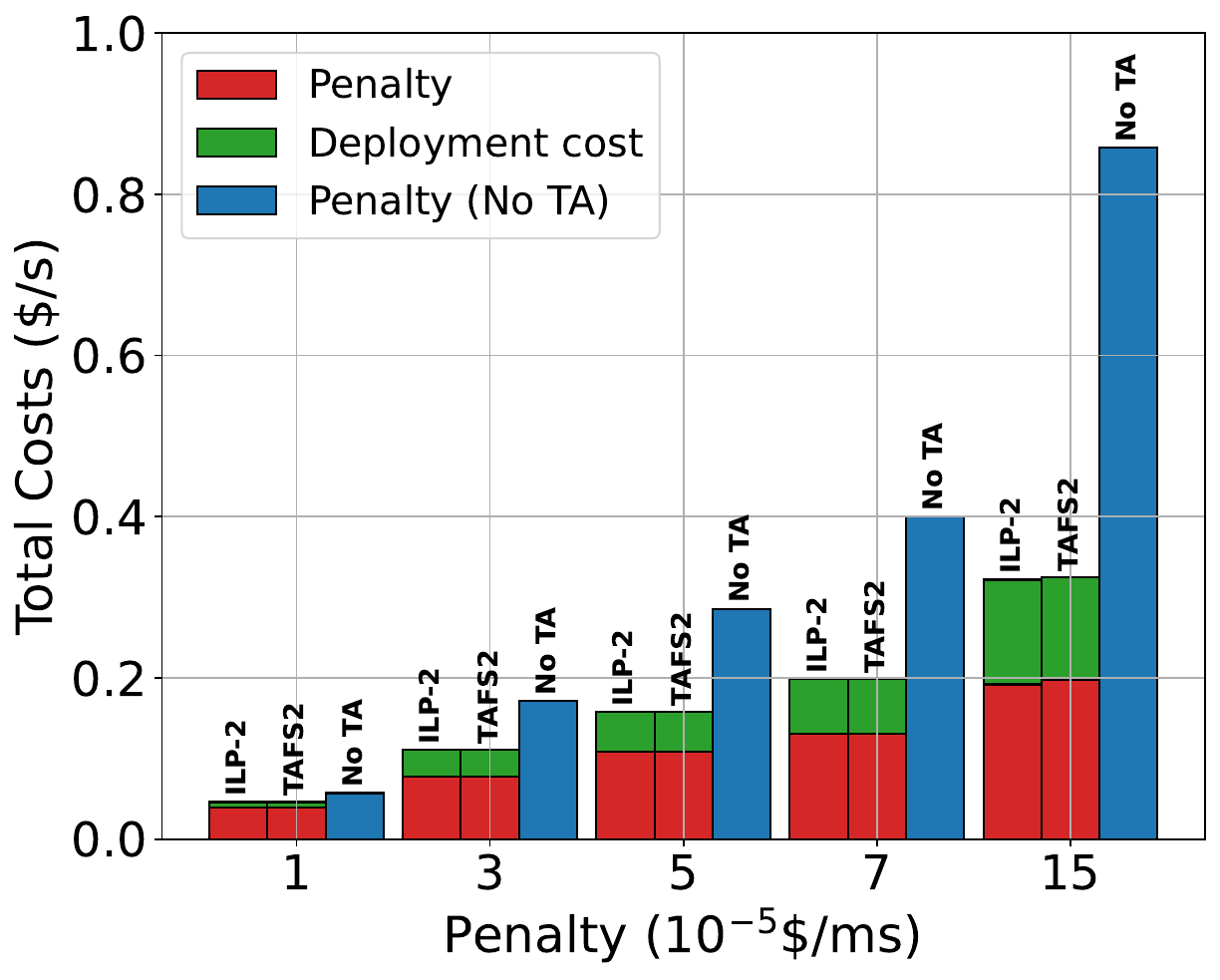}\label{fig:costs_abilene_heuristic2}}
    \hspace{15pt}
    \subfloat[Total costs with and without TAs for the studied topologies (Penalty:~$5*10^{-5}$\$/ms)]{\includegraphics[width=0.30\textwidth]{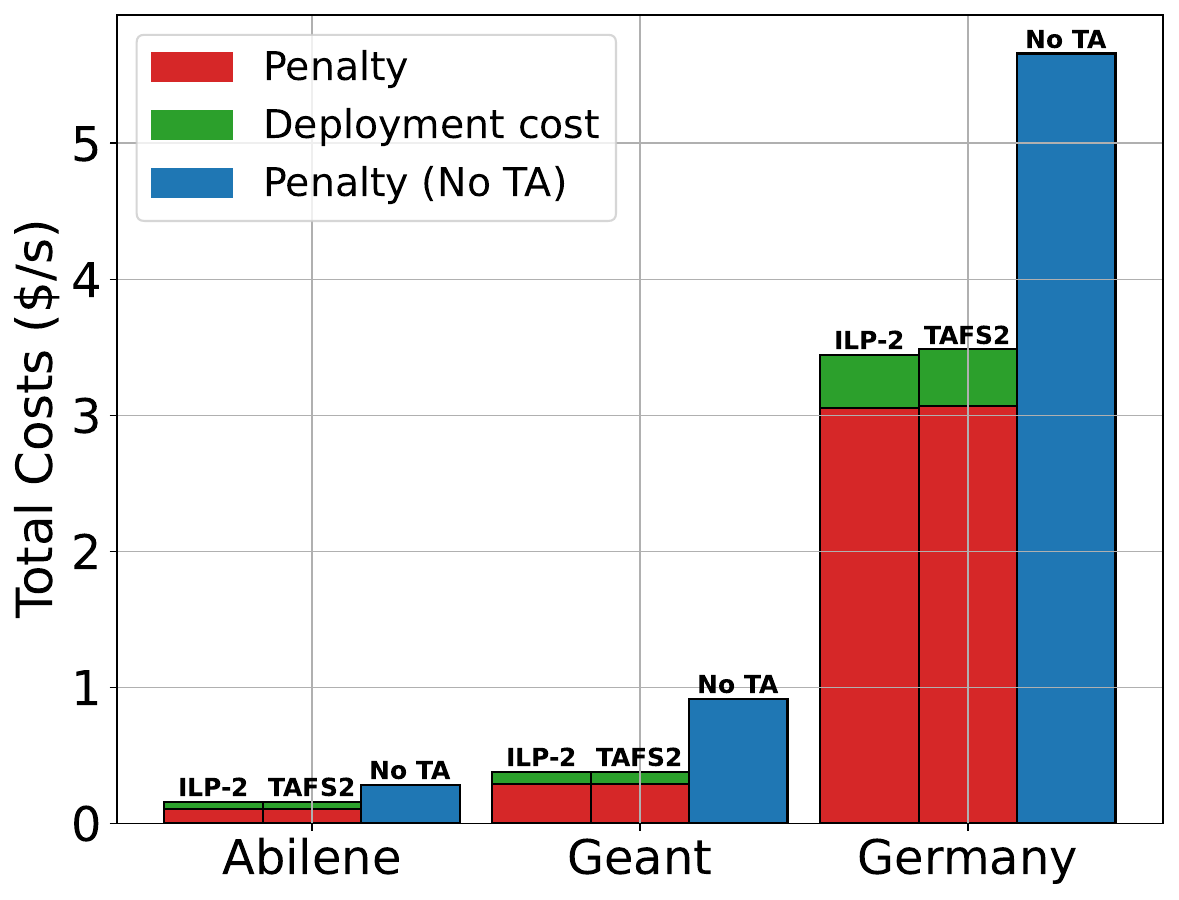}\label{fig:costs_topologies_heuristic2}}
    \caption{Results for TAFS2.}
    \label{fig:overall-opt-vs-heuristic-2}
    \end{center}
\end{figure*}

In the case of TAFS2 (objective 2) compared to ILP2, we~can see in Figure~\ref{fig:overall-opt-vs-heuristic-2} how TAFS2 is able to find the optimal solution in all tests, except when the penalty value increases to $15*10^{-5}$ \$/ms, where the total cost increases by 0.98\% (Figure~\ref{fig:costs_abilene_heuristic2}). 
On the other hand, in the scalability analysis (Figure~\ref{fig:costs_topologies_heuristic2}), we obtain similar results: TAFS2 is able to find the optimal solution in all tests except in Germany (when the problem size is large), where the total cost increases by 1.2\%.

Finally, in terms of execution time, we can see in Table~\ref{tab:execution_times} that TAFS2 achieves these results by spending 9.76, 19.34, 8.53 less computation time in Abilene, Geant and Germany, respectively.

\begin{table} [htb]
    \centering
    \caption{Execution times}
    \label{tab:execution_times}
    \begin{tabular}{cccc}
    \hline
         & Abilene & Geant & Germany\\
         \hline
        ILP - Objective 1 & 7.22 s & 74.47 s & 5838.6 s\\
        ILP - Objective 2 & 1.66 s & 13.15 s & 171.33 s\\
        TAFS - Objective 1 & 0.008 s & 0.033 s & 1.57 s\\
        TAFS - Objective 2 & 0.17 s & 0.68 s & 20.08 s\\
        \hline
    \end{tabular}
\end{table}

\section{Related Work}\label{sec:RelatedWork}

In recent decades, numerous studies have focused on improving the performance of TCP and transport protocols in general~\cite{zhani2020flexngia, Zhani-WMNC22, Luglio2004, wang2017sdudp, chen2019sdatp, chen2019sdn, wan2002psfq, TCPSplicing2000, Rosu2002, 10024836, pokhrel2020improving, BRUHN2023109609}. For example, Rosuetal.\cite{Rosu2002} proposed splitting TCP connections between clients and servers by using proxy servers. They introduced a TCP Splicing service that optimizes communication within split-connection proxies by removing the need for duplicating packets between the kernel and application buffers. Wang et al.\cite{wang2017sdudp} suggested using edge switches to convert TCP to UDP, delegating packet retransmissions to intermediate switches. Similarly, Wan~et~al.~\cite{wan2002psfq} aimed to reduce retransmission delays in~wireless networks by introducing a transport protocol based on hop-by-hop transmission, where each hop detects packet loss by~identifying missing sequence numbers.

Chen et al.~\cite{chen2019sdn, chen2019sdatp} proposed a transport protocol leveraging the control plane of software-defined networks to establish connections, and the data plane to handle caching and retransmissions. Unlike these approaches, which require modifications to TCP, the TA function presented in \cite{Zhani-WMNC22} supports TCP without needing alterations. This function is deployed within the network to cache and retransmit TCP packets, all while keeping the TCP endpoints unaware of its existence. However, the work in \cite{Zhani-WMNC22} does not address the placement of~the~TA~function, nor does it offer a mathematical model to evaluate the packet delivery delays when the TA is utilized.

In addition, our previous work \cite{LuisNOMS2023} proposed placing TA in network segments identified as overloaded using machine learning techniques. Nevertheless, it did not take into account factors such as packet loss or propagation delays to optimize the TA's placement, nor did it evaluate their impact on delivery delays.

To the best of our knowledge, this is the first study to model the impact of TA placement on packet delivery delays using a mathematical approach, allowing us to quantify the performance benefits for TCP. Furthermore, this work is the first to propose efficient algorithms for TA placement that consider packet delivery delay requirements.
\section{Conclusion} \label{sec:conclusion}



\color{black}
While prior work has demonstrated the benefits of TAs and~studied their placement along fixed routes, this paper addressed a more general and practical setting where both routing and TA placement decisions are made jointly in order to minimize average packet delivery delay for TCP packets.

To this end, we formulated the joint flow routing and TA placement problem under two distinct scenarios: (i)~a~best-effort setting aiming only to minimize average packet delivery delay, and (ii) a QoS-constrained setting where flows are subject to SLA requirements and TA deployment incurs costs. We~proposed Integer Linear Programming formulations for~both scenarios and developed an efficient heuristic to handle larger instances. Our~extensive simulations confirm that~the~heuristic achieves near-optimal performance: it significantly reduces average delivery delays in best-effort networks and ensures SLA satisfaction with minimal deployment costs in~QoS-constrained networks.

Overall, this work highlights the potential of TAs as~a~powerful tool to enhance transport performance and demonstrates the necessity of integrating their placement into routing decisions. Future research directions include extending the framework to dynamic traffic matrices, considering energy-aware and multi-resource optimization objectives, and developing online algorithms that can adapt TA placement and~routing decisions in real time.
\color{black}

\section*{Acknowledgements}

This work has been partially funded by the Ministry of Science, Innovation and Universities (project PID2021-124054OB-C31) and by the European Union, European Regional Development Fund, and the Regional Government of Extremadura. Managing Authority: Ministry of Finance. Aid file number: GR24099.

\bibliographystyle{elsarticle-num} 
\bibliography{bibliography}

@INPROCEEDINGS{JaimeTAPlacementNOMS2024,
  author={Galán-Jiménez, Jaime and Zhani, Mohamed Faten and Jesús Martín León, Luis and Kaippallimalil, John},
  booktitle={IEEE Network Operations and Management Symposium (NOMS)}, 
  title={{Transport Assistants to Enhance TCP Performance: Analysis of the Packet Delivery Delay}}, 
  year={2024},
  volume={},
  number={},
  pages={1-5},
  doi={10.1109/NOMS59830.2024.10575037}}

@Misc{Gurobipy,
  title = {Gurobipy Optimizer},
  note = {https://pypi.org/project/gurobipy/, last accessed: 2025-09-10},
}

@INPROCEEDINGS{LuisNOMS2023,
  author={León, Luis Jesús Martín and Herrera, Juan Luis and Berrocal, Javier and Galán-Jiménez, Jaime},
  booktitle={NOMS 2023-2023 IEEE/IFIP Network Operations and Management Symposium}, 
  title={Logistic Regression-based Solution to Predict the Transport Assistant Placement in SDN networks}, 
  year={2023},
  volume={},
  number={},
  pages={1-5},
  doi={10.1109/NOMS56928.2023.10154303}}

@ARTICLE{Cao2023,
  author={Cao, Jie and Zhu, Xu and Sun, Sumei and Wei, Zhongxiang and Jiang, Yufei and Wang, Jingjing and Lau, Vincent K.N.},
  journal={IEEE Wireless Communications}, 
  title={{Toward Industrial Metaverse: Age of Information, Latency and Reliability of Short-Packet Transmission in 6G}}, 
  year={2023},
  volume={30},
  number={2},
  pages={40-47},
  doi={10.1109/MWC.2001.2200396}}

@INPROCEEDINGS{Yahyaoui-NOMS22,
  author={Yahyaoui, Haythem and Majdoub, Melek and Zhani, Mohamed Faten and Aloqaily, Moayad},
  booktitle={IEEE/IFIP Network Operations and Management Symposium (NOMS)}, 
  title={{On Minimizing TCP Retransmission Delay in Softwarized Networks}}, 
  year={2022},
  pages={1-6},
  doi={10.1109/NOMS54207.2022.9789762}}

@misc{rfc793,
    series =    {Request for Comments},
    number =    793,
    howpublished =  {RFC 793},
    publisher = {RFC Editor},
    doi =       {10.17487/RFC0793},
        author =    {},
    title =     {{Transmission Control Protocol}},
    pagetotal = 91,
    year =      1981,
    month =     sep,
}

@INPROCEEDINGS{Hamad2023,
  author={Hamad, Diyar Jamal and Yalda, Khirota Gorgees and Tapus, Nicolae and Okumus, Ibrahim Taner},
  booktitle={International Conference on Control Systems and Computer Science (CSCS)}, 
  title={{Network Management devices in an SDN environment}}, 
  year={2023},
  pages={86-91},
  doi={10.1109/CSCS59211.2023.00023}}

@article{Yahyaoui2023,
author = {Yahyaoui, Haythem and Zhani, Mohamed Faten and Bouachir, Ouns and Aloqaily, Moayad},
title = {On minimizing flow monitoring costs in large-scale software-defined network networks},
journal = {International Journal of Network Management (IJNM)},
volume = {33},
number = {2},
pages = {e2220},
doi = {https://doi.org/10.1002/nem.2220},
year = {2023}
}

@INPROCEEDINGS{Tao2023,
  author={Tao, Hou-Yeh and Huang, Chih-Kai and Shen, Shan-Hsiang},
  booktitle={IEEE Symposium on Computers and Communications (ISCC)}, 
  title={{A Low-overhead Network Monitoring for SDN-Based Edge Computing}}, 
  year={2023},
  volume={},
  number={},
  pages={600-606},
  doi={10.1109/ISCC58397.2023.10218002}}

@article{zhani2020flexngia,
  author    = {Mohamed Faten Zhani and Hesham ElBakoury},
  title     = {{FlexNGIA: A Flexible Internet Architecture for the Next-Generation Tactile Internet}},
  journal   = {Journal of Network and Systems Management},
  year      = {2020},
  volume    = {28},
  number    = {4},
  pages     = {751--795},
  doi       = {10.1007/s10922-020-09525-0},
  url       = {https://doi.org/10.1007/s10922-020-09525-0},
  issn      = {1573-7705}
}

@INPROCEEDINGS{Zhani-WMNC22,
  author={Zhani, Mohamed Faten and Boughamoura, Rihab and Yahyaoui, Haythem and Kaippallimalil, John and Kiani, Abbas},
  booktitle={IFIP Wireless and Mobile Networking Conference (WMNC)}, 
  title={{Oblivious TCP Support - A Virtual Network Function to Speed Up TCP in Wireless Environments}}, 
  year={2022},
  pages={75-79},
  doi={10.23919/WMNC56391.2022.9954287}}

@inproceedings{Rosu2002,
author = {Ro\c{s}u, Marcel-Cundefinedtundefinedlin and Ro\c{s}u, Daniela},
title = {{An evaluation of TCP splice benefits in web proxy servers}},
year = {2002},
isbn = {1581134495},
publisher = {Association for Computing Machinery},
address = {New York, NY, USA},
url = {https://doi.org/10.1145/511446.511449},
doi = {10.1145/511446.511449},
booktitle = {International Conference on World Wide Web},
pages = {13–24},
numpages = {12},
location = {Honolulu, Hawaii, USA},
series = {WWW '02}
}

@article{moufakirITU2022,
  title={{SFCaaS: Service function chains as~a~service in~NFV environments}},
  author={Tarik Moufakir and Mohamed Faten Zhani and Abdelouahed Gherbi and  Moayad Aloqaily and Nadir Ghrada},
  journal={ITU Journal on Future and Evolving Technologies},
  volume={3},
  number={3},
  pages={679--692},
  year={2022},
	 publisher={ITU},
  doi={https://doi.org/10.52953/ZPDB8065}
}

@INPROCEEDINGS{LuizelliIM2015,
  author={Luizelli, Marcelo Caggiani and Bays, Leonardo Richter and Buriol, Luciana Salete and Barcellos, Marinho Pilla and Gaspary, Luciano Paschoal},
  booktitle={IFIP/IEEE International Symposium on Integrated Network Management (IM)}, 
  title={Piecing together the NFV provisioning puzzle: Efficient placement and chaining of virtual network functions}, 
  year={2015},
  volume={},
  number={},
  pages={98-106},
  doi={10.1109/INM.2015.7140281}}

@ARTICLE{TCPSplicing2000,
  author={Spatscheck, O. and Hansen, J.S. and Hartman, J.H. and Peterson, L.L.},
  journal={IEEE/ACM Transactions on Networking}, 
  title={{Optimizing TCP forwarder performance}}, 
  year={2000},
  volume={8},
  number={2},
  pages={146-157},
  doi={10.1109/90.842138}}

@ARTICLE{chen2019sdatp,
  author={Chen, Jiayin and Ye, Qiang and Quan, Wei and Yan, Si and Do, Phu Thinh and Zhuang, Weihua and Shen, Xuemin Sherman and Li, Xu and Rao, Jaya},
  journal={IEEE Network}, 
  title={{SDATP: An SDN-Based Adaptive Transmission Protocol for Time-Critical Services}}, 
  year={2020},
  volume={34},
  number={3},
  pages={154-162},
  doi={10.1109/MNET.001.1900333}}

@ARTICLE{wang2017sdudp,
  author={Wang, Ming-Hung and Chen, Lung-Wen and Chi, Po-Wen and Lei, Chin-Laung},
  journal={IEEE Access}, 
  title={{SDUDP: A Reliable UDP-Based Transmission Protocol Over SDN}}, 
  year={2017},
  volume={5},
  number={},
  pages={5904-5916},
  doi={10.1109/ACCESS.2017.2693376}}

@INPROCEEDINGS{chen2019sdn,
  author={Chen, Jiayin and Yan, Si and Ye, Qiang and Quan, Wei and Do, Phu Thinh and Zhuang, Weihua and Shen, Xuemin Sherman and Li, Xu and Rao, Jaya},
  booktitle={IEEE International Conference on Communications (ICC)}, 
  title={{An SDN-Based Transmission Protocol with In-Path Packet Caching and Retransmission}}, 
  year={2019},
  volume={},
  number={},
  pages={1-6},
  doi={10.1109/ICC.2019.8761782}}

@inproceedings{wan2002psfq,
author = {Wan, Chieh-Yih and Campbell, Andrew T. and Krishnamurthy, Lakshman},
title = {{PSFQ: a reliable transport protocol for wireless sensor networks}},
year = {2002},
isbn = {1581135890},
publisher = {Association for Computing Machinery},
address = {New York, NY, USA},
doi = {10.1145/570738.570740},
booktitle = {ACM International Workshop on Wireless Sensor Networks and Applications},
pages = {1–11},
numpages = {11},
location = {Atlanta, Georgia, USA},
series = {WSNA '02}
}

@article{Luglio2004,
author = {Luglio, Michele and Roseti, Cesare and Gerla, M},
year = {2004},
month = {06},
pages = {1-9},
title = {{The impact of efficient flow control and OS features on TCP performance over satellite links}},
volume = {3},
journal = {ASSI Satellite Communication Letter (Sat-Comm Letter)}
}

@article{SNDlib2010,
author = {Orlowski, S. and Wessäly, R. and Pióro, M. and Tomaszewski, A.},
title = {SNDlib 1.0—Survivable Network Design Library},
journal = {Networks},
volume = {55},
number = {3},
pages = {276-286},
doi = {https://doi.org/10.1002/net.20371},
url = {https://onlinelibrary.wiley.com/doi/abs/10.1002/net.20371},
eprint = {https://onlinelibrary.wiley.com/doi/pdf/10.1002/net.20371},
year = {2010}
}

@ARTICLE{10024836,
  author={Mishra, Tapas Kumar and Sahoo, Kshira Sagar and Bilal, Muhammad and Shah, Sayed Chhattan and Mishra, Manas Kumar},
  journal={IEEE Access}, 
  title={{Adaptive Congestion Control Mechanism to Enhance TCP Performance in Cooperative IoV}}, 
  year={2023},
  volume={11},
  number={},
  pages={9000-9013},
  doi={10.1109/ACCESS.2023.3239302}}

@ARTICLE{pokhrel2020improving,
  author={Pokhrel, Shiva Raj and Choi, Jinho},
  journal={IEEE Transactions on Vehicular Technology}, 
  title={{Improving TCP Performance Over WiFi for Internet of Vehicles: A Federated Learning Approach}}, 
  year={2020},
  volume={69},
  number={6},
  pages={6798-6802},
  doi={10.1109/TVT.2020.2984369}}

@ARTICLE{Liu2023,
  author={Liu, Yan and Deng, Yansha and Nallanathan, Arumugam and Yuan, Jinhong},
  journal={IEEE Wireless Communications}, 
  title={{Machine Learning for 6G Enhanced Ultra-Reliable and Low-Latency Services}}, 
  year={2023},
  volume={30},
  number={2},
  pages={48-54},
  doi={10.1109/MWC.006.2200407}}

@article{BRUHN2023109609,
title = {{Performance and improvements of TCP CUBIC in low-delay cellular networks}},
journal = {Computer Networks},
volume = {224},
pages = {109609},
year = {2023},
issn = {1389-1286},
doi = {https://doi.org/10.1016/j.comnet.2023.109609},
author = {Philipp Bruhn and Mirja Kühlewind and Maciej Muehleisen}
}

@book{cormen2009introduction,
  title={Introduction to Algorithms},
  author={Cormen, Thomas H. and Leiserson, Charles E. and Rivest, Ronald L. and Stein, Clifford},
  year={2009},
  publisher={MIT press},
  edition={3rd},
  address={Cambridge, MA},
  isbn={978-0-262-03384-8}
}

@article{yen1971finding,
author = {Yen, Jin Y.},
title = {Finding the K Shortest Loopless Paths in a Network},
journal = {Management Science},
volume = {17},
number = {11},
pages = {712-716},
year = {1971},
doi = {10.1287/mnsc.17.11.712},
URL = { https://doi.org/10.1287/mnsc.17.11.712},
eprint = { https://doi.org/10.1287/mnsc.17.11.712}
}

@article{eppstein1998finding,
author = {Eppstein, David},
title = {Finding the k Shortest Paths},
journal = {SIAM Journal on Computing},
volume = {28},
number = {2},
pages = {652-673},
year = {1998},
doi = {10.1137/S0097539795290477},
URL = {https://doi.org/10.1137/S0097539795290477},
eprint = {https://doi.org/10.1137/S0097539795290477}
}

\end{document}